\documentclass[aps,superscriptaddress,showkeys,preprintnumbers,showpacs,nofootinbib]{revtex4}
\usepackage{graphicx,wrapfig,amssymb}
\newcommand{\be}{\begin{equation}}
\newcommand{\ee}{\end{equation}}
\newcommand{\ba}{\begin{eqnarray}}
\newcommand{\ea}{\end{eqnarray}}
\newcommand{\Mn}{M_{\rm N}}
\newcommand{\pE}{p_{\mbox{\tiny E}}}
\newcommand{\sigmaPiN}{\sigma_{\pi\rm N}}

\newcommand{\fslash}[1] {{\not\! #1\,}}

\newcommand{\di}{ {\rm d} }
\newcommand{\la}{\langle}
\newcommand{\ra}{\rangle}

\newcommand{\doublesum}[2]
	{\renewcommand{\arraystretch}{0.7} \begin{array}{c}
	{\displaystyle\sum} \\
	    \scriptstyle {#1} \\ \scriptstyle {#2} \end{array}}
\begin{document}
%===================  TITLE, AUTHORS, AFFILIATIONS ===================
\newcommand*{\Bochum}{Institut f{\"u}r Theoretische Physik II, 
	Ruhr-Universit{\"a}t Bochum, D-44780 Bochum, Germany}\affiliation{\Bochum}
\newcommand*{\Porto}{Faculdade de Engenharia da Universidade do Porto,
	P-4000 Porto, Portugal}\affiliation{\Porto}
\newcommand*{\Coimbra}{Centro de F\'isica Computacional, Universidade de
	Coimbra, P-3000 Coimbra, Portugal}
\title{	The twist-3 parton distribution function \boldmath $e^a(x)$ 
	in large-$N_c$ chiral theory}
\author{C.~Cebulla}\affiliation{\Bochum}
\author{J.~Ossmann}\affiliation{\Bochum}
\author{P.~Schweitzer}\affiliation{\Bochum}
\author{D.~Urbano}\affiliation{\Porto}\affiliation{\Coimbra}
\date{September 2007}
%===================  PREPRINT NUMBER, JOURNAL =======================
\preprint{\tt RUB-TPII-07-07}
%===================  ABSTRACT =======================================
\begin{abstract}
The chirally-odd twist-3 parton distribution function $e^a(x)$ of the nucleon 
is studied in the large-$N_c$ limit in the framework of the chiral quark-soliton model.
It is demonstrated that in spite of properties not shared by other distribution 
functions, namely the appearance of a $\delta(x)$-singularity and quadratic 
divergences in $e^a(x)$, an equally reliable calculation is possible. 
Among the most remarkable results obtained in this work is the fact that the 
coefficient of the $\delta(x)$-singularity can be computed exactly in this model,
avoiding involved numerics.
Our results complete existing studies in literature. 
\end{abstract}
\pacs{
  12.39.Fe, % Chiral Lagrangians
  11.15.Pg, % Expansions for large numbers of components (e.g., 1/Nc expansions)
  12.38.Lg, % Other nonperturbative calculations
  13.60.Hb} % Total and inclusive cross sections (including deep-inelastic processes)
% 12.38.Gc  % Lattice QCD calculations
% 12.39.Dc  % Skyrmions
% 12.39.Ki  % Relativistic quark model
% 14.20.Dh  % Protons and neutrons
% 13.88.+e  % Polarization in interactions and scattering
% 13.85.Ni  % Inclusive production with identified hadrons
\keywords{
chirally odd twist-3 parton distribution, 
large $N_c$ limit, 
chiral soliton model of the nucleon}

\maketitle
%====== SECTION 1: INTRODUCTION ======================================
\section{Introduction}
\label{Sec-1:introduction}

The twist-3 parton distribution function $e^a(x)$ \cite{Jaffe:1991ra} was so far 
subject to modest interest in literature, in spite of its remarkable theoretical 
properties, because of its chirally odd nature which makes it difficult to access 
in experiment. The probably most striking theoretical property is the existence 
of a $\delta$-function-type singularity at $x=0$ in $e^a(x)$ which follows from 
the QCD equations of motion \cite{Balitsky:1996uh,Belitsky:1997ay,Efremov:2002qh}.

Since it is chirally odd $e^a(x)$ can contribute to an observable only in connection
with another chirally odd object.
For example,
$e^a(x)$ and the 
chirally odd Collins fragmentation function $H_1^\perp$ \cite{Collins:1992kk} 
contribute to an azimuthal asymmetry in semi-inclusive deeply inelastic scattering 
of longitudinally polarized electrons off unpolarized nucleons 
\cite{Levelt:1994np,Mulders:1995dh} which was measured 
\cite{Airapetian:1999tv,Avakian:2003pk,Airapetian:2006rx} and used 
to extract first information on $e^a(x)$ \cite{Efremov:2002ut}.
Later it became clear that $e^a(x)$ and $H_1^\perp$ are not the 
only contribution to this observable 
\cite{Afanasev:2003ze,Yuan:2003gu,Bacchetta:2004zf,Metz:2004je,Bacchetta:2006tn}.
Studies of two-hadron production in  semi-inclusive deeply inelastic scattering 
could provide a more direct and easier access to $e^a(x)$ \cite{Bacchetta:2003vn}.

In experiment the $\delta(x)$-singularity could be observed only indirectly, namely 
as a discrepancy for the first moment of $e^a(x)$ between the theoretical result,
which includes the point $x=0$, and an experimental result where only $x>0$ contribute 
\cite{Efremov:2002qh}.
A ``direct observation'' of the $\delta(x)$-contribution, however, is possible 
in models. 

The appearance of a $\delta(x)$-singularity makes $e^a(x)$ a particularly
interesting but, at the same time, also demanding object for model studies. 

While there is no such singularity in bag \cite{Jaffe:1991ra,Signal:1996ct} or 
spectator \cite{Jakob:1997wg} models, a $\delta(x)$-contribution in $e^a(x)$ 
was found in $(1+1)$-dimensional toy model or perturbative one-loop 
calculations \cite{Burkardt:1995ts,Burkardt:2001iy}, and in the 
chiral quark-soliton model ($\chi$QSM) 
\cite{Schweitzer:2003uy,Wakamatsu:2003uu,Ohnishi:2003mf}.
The latter is the ground for the present study. 

The existence of a $\delta(x)$-contribution in $e^a(x)$ in the $\chi$QSM 
was proven independently in \cite{Schweitzer:2003uy,Wakamatsu:2003uu}.
Moreover in \cite{Schweitzer:2003uy} a first approximate calculation 
of $e^a(x)$ was presented and a method was suggested, how to compute 
in practice in this model a parton distribution function in which a 
$\delta(x)$-singularity appears. This method was explored to calculate
$e^a(x)$ exactly in the $\chi$QSM in \cite{Ohnishi:2003mf}.

The calculation of $e^a(x)$ in the $\chi$QSM is demanding 
not only because of the appearance of the $\delta(x)$-contribution.
A further complication is due to the fact that $e^a(x)$ is quadratically UV-divergent,
in contrast to the parton distribution functions studied so far which are
UV-finite or at worst logarithmically divergent.
For these reasons it is worthwhile to present an independent exact calculation 
of $e^a(x)$ in the $\chi$QSM. This is the purpose of the present study.

In order to supplement and complete previous works 
\cite{Schweitzer:2003uy,Wakamatsu:2003uu,Ohnishi:2003mf}
we shall devote particular effort to the demonstration that 
the involved numerics in this calculation is well under control.
The tool we shall use for that is the so-called gradient expansion.
This work, however,  goes beyond a mere reexamination and confirmation of the 
previous studies \cite{Schweitzer:2003uy,Ohnishi:2003mf}. 
In fact, among the new insights, the most remarkable one   is the observation 
that the result for the coefficient of the $\delta(x)$-term in $e^a(x)$,
obtained by means of the gradient expansion in \cite{Schweitzer:2003uy}, 
is {\sl exact}.

This note is organized as follows.
In Sec.~\ref{Sec-2:e-theory} we review the general properties of $e^a(x)$.
In Sec.~\ref{Sec-3:model-and-e}	we introduce the $\chi$QSM, and discuss how 
$e^a(x)$ is described in this model.
The Secs.~\ref{Sec-4:(eu+ed)},~\ref{Sec-5:(eu+ed)-and-delta} and \ref{Sec-6:x(eu+ed)} 
are devoted to the singlet  flavour combination which is leading in the large-$N_c$ 
limit, and we discuss in detail respectively its singular and regular parts,
and the origin of the $\delta(x)$-singularity.
In Sec.~\ref{Sec-7:(eu-ed)} we study the non-singlet flavour combination 
which is the subleading structure in the large-$N_c$ expansion.
In Sec.~\ref{Sec-8:results-sum-rules} we summarize and discuss our results which we 
compare to previous studies in the $\chi$QSM and other models of $e^a(x)$
in Secs.~\ref{Sec-9:previous-in-model} and \ref{Sec-10:compare-to-other-models}, 
respectively. Sec.~\ref{Sec-11:conclusions} contains the conclusions.

%\newpage
%====== SECTION 2: e(x) in theory ====================================
\section{\boldmath The distribution function $e^a(x)$}
\label{Sec-2:e-theory}

Let us first recall some general theoretical properties of $e^a(x)$,
a more extensive review can be found in Ref.~\cite{Efremov:2002qh}.
The twist-3 distribution functions $e^q(x)$ for quarks 
of flavour $q$ and $e^{\bar q}(x)$ for antiquarks of flavour $\bar q$ 
are defined as \cite{Jaffe:1991ra}
\ba\label{Eq:def-e}
	e^q(x) = \frac{1}{2\Mn} \int\!\frac{\di\lambda}{2\pi}\,e^{i\lambda x}\,
	\la N|\,\bar{\psi}_q(0)\,[0,\lambda n]\,\psi_q(\lambda n)\,|N\ra
	\;,\;\;\;
	e^{\bar q}(x) = e^q(-x) \;,
	\ea
with the light-like vector $n$ and $[0,\lambda n]$ denoting the gauge-link. 
The dependence on the renormalization scale $\mu$, which we do not indicate in 
(\ref{Eq:def-e}) for brevity, was studied in \cite{Balitsky:1996uh,Belitsky:1997ay}.

The equations of motion of QCD allow to decompose $e^q(x)$ 
as \cite{Balitsky:1996uh,Belitsky:1997ay,Efremov:2002qh}
(in this work we shall consider $e^q(x)$ in the chiral limit, 
and limit ourselves to only indicate current quark mass effects)
\be\label{e-decomposition}
	e^q(x) =\frac{\delta(x)}{2\Mn}\,\la N|\bar\psi_q(0)\psi_q(0)|N\ra 
	       + e^q_{\rm tw3}(x) + {\cal O}\left(\frac{m_q}{\Mn}\right).\ee
The contribution $e^q_{\rm tw3}(x)$ in 
Eq.~(\ref{e-decomposition}) is a quark-gluon-quark correlation function, i.e. 
the actual ``pure'' twist-3 (``interaction dependent'') contribution to 
$e^a(x)$, and has a partonic interpretation as an interference between 
scattering from a coherent quark-gluon pair and from a single quark 
\cite{Jaffe:1991ra}. Its first two moments vanish. Consequently, the 
first moment of $e^q(x)$ is saturated by the $\delta(x)$-contribution and 
the second is due to quark mass effects only 
\ba\label{e-1moment}
	\int_{-1}^1\di x\;e^q(x) &=& 
	\frac{1}{2\Mn}\la N|\,\bar{\psi}_q(0)\psi_q(0)\,|N\ra\;,\\
\label{e-2moment}
	\int_{-1}^1\di x\;x\,e^q(x) &=& \frac{m_q}{\Mn}\; N_q \;. \ea
In Eq.~(\ref{e-2moment}) $N_q$ denotes the number of the respective valence quarks 
(for proton $N_u=2$ and $N_d=1$). 

The first moment (\ref{e-1moment}) of the flavour singlet combination is related 
to the pion-nucleon sigma-term $\sigmaPiN$ as
\be\label{e-1moment-a}
	\int_{-1}^1\di x\;(e^u+e^d)(x)=
	\frac{\sigmaPiN}{m}\equiv\frac{\sigma(t)}{m}\biggl|_{t=0}
	\approx (5-15)\;,
\ee
where $m=\frac12\,(m_u+m_d)$ and $\sigma(t)$ denotes the scalar isoscalar nucleon 
form-factor which is known only at the Cheng-Dashen point $t=2 m_\pi^2$ where it 
is related by low-energy theorems to pion-nucleon scattering amplitudes. Analyses 
yield $\sigma(2m_\pi^2) = (56-88)\,{\rm MeV}$ \cite{Koch:1982pu,Pavan:2001wz}.
The difference $\sigma(2m_\pi^2)-\sigma(0)=14\,{\rm MeV}$ was obtained from a 
dispersion relation analysis \cite{Gasser:1990ce} and a similar result was 
found in chiral perturbation theory calculations \cite{Becher:1999he}. Thus 
$\sigmaPiN \approx (42-74)\;{\rm MeV}$ and with $m = (7\pm2)\,{\rm MeV}$
\cite{Gasser:1982ap} one obtains the result quoted in Eq.~(\ref{e-1moment-a})
which refers to the scale $\mu=1\,{\rm GeV}$.

The flavour non-singlet combination satisfies an equally interesting sum rule, namely
\be\label{e-1moment-c}
	\int_{-1}^1\di x\;(e^u-e^d)(x) = \frac{(M_n-M_p)_{\rm hadr}}{m_d-m_u}
	\approx 0.4\;.
\ee
Here $(M_n-M_p)_{\rm hadr}\approx 2\,{\rm MeV}$  \cite{Gasser:1982ap} denotes the 
hadronic mass difference between neutron and proton, i.e.\  the mass difference 
in the absence of electro-weak interactions. With $m_d-m_u\approx 5\,{\rm MeV}$
\cite{Gasser:1982ap} one obtains the result quoted in Eq.~(\ref{e-1moment-c})
at a scale of $\mu=1\,{\rm GeV}$.
It is important to keep in mind that the numbers in 
Eqs.~(\ref{e-1moment-a},~\ref{e-1moment-c}) are not due to the ``valence'' 
structure of $e^q(x)$, but solely due to the $\delta(x)$-contributions.

In the large-$N_c$ limit the behaviour of the different flavour combinations
of $e^a(x)$ is as follows \cite{Efremov:2002qh,Schweitzer:2003uy}
\ba\label{Eq:large-Nc}
	(e^u+e^d)(x) &=& N_c^2\:d_+(N_c x) \;,\nonumber\\
	(e^u-e^d)(x) &=& N_c\,\:d_-(N_c x) \;,\ea  
where $d_\pm(y)$ are stable functions in the limit $N_c\to\infty$ for fixed 
arguments $y=N_cx$. This implies the following hierarchy 
\be\label{Eq:large-Nc-2}
	|(e^u+e^d)(x)| \gg |(e^u-e^d)(x)|
\ee
in the large-$N_c$ limit.
Notice that the twist-2 distribution function $f_1^a(x)$ 
exhibits the analog flavour dependence. Noteworthy, $e^q(x)$ and $f_1^q(x)$ 
become equal in the non-relativistic limit \cite{Efremov:2002qh,Schweitzer:2003uy}.

%\newpage
%====== SECTION 3: CHIRAL QUARK-SOLITON MODEL AND e(x) ===============
\section{
Chiral quark soliton model, its applications, limitations, and \boldmath $e^a(x)$} 
\label{Sec-3:model-and-e}

The effective theory underlying the chiral quark soliton model ($\chi$QSM) 
\cite{Diakonov:1987ty}  was derived from the instanton model of the QCD vacuum 
\cite{Diakonov:1983hh,Diakonov:1995qy}, and is given by the partition function 
\cite{Diakonov:1986yh}
\be\label{eff-theory}
	Z_{\rm eff} = \int\!{\cal D}\psi\,{\cal D}\bar{\psi}\,{\cal D}U\,
	\exp\biggl[i\int\!\!\di^4x\;\bar{\psi}\,
	(i\fslash{\partial}-M\,U^{\gamma_5}-m)\psi\biggr],\ee
where $U^{\gamma_5}=\exp(i\gamma_5\tau^a\pi^a)$. $M$ denotes the dynamical quark mass 
due to the spontaneous breakdown of chiral symmetry, and $m$ is the current quark mass 
responsible for explicit chiral symmetry breaking effects.
The effective theory (\ref{eff-theory}) contains the Wess-Zumino term  and the 
four-derivative Gasser-Leutwyler terms with correct coefficients. 
This effective theory provides an approximation to the dynamics of light quarks valid 
at low energies below 
\be\label{scale}
	\rho_{\rm av}^{-1} \approx 600\,{\rm MeV}, \ee
where $\rho_{\rm av}$ is the average instanton size.
Corrections to this picture are of the order $(M\rho_{\rm av})^2\sim 30\%$.
It is important to remark that $(M\rho_{\rm av})^2$ is proportional to the 
parametrically small instanton packing fraction
\be\label{Eq:packing-fraction}
	(M\rho_{\rm av})^2 \propto 
	\biggl(\frac{\rho_{\rm av}}{R_{\rm av}}\biggr)^{\!4} \ll 1 \;,\ee
where $R_{\rm av}$ denotes the average distance between instantons. The numerical
smallness of the parameter $\rho_{\rm av}/R_{\rm av}\sim\frac13$ played an important 
role in the derivation of the effective theory (\ref{eff-theory}) from the instanton 
vacuum model \cite{Diakonov:1983hh}.

Let us briefly recall how the effective theory (\ref{eff-theory}) describes nucleons
\cite{Diakonov:1987ty}. In the leading order of the large-$N_c$ limit the pion field 
is static and the soliton energy $E_{\rm sol}$ is a functional of this field given by
\be\label{soliton-energy}
	E_{\rm sol}[U] = N_c \biggl(E_{\rm lev}+
	\sum\limits_{E_n<0}(E_n-E_{n_0})\biggr)\biggr|_{\rm reg} \;. \ee 
The $E_n$ in (\ref{soliton-energy}) are the eigen-energies of the one-particle 
Hamiltonian
\be\label{Hamiltonian}
	\hat{H}|n\ra=E_n |n\ra \;,\;\;
	\hat{H}=-i\gamma^0\gamma^k\partial_k+\gamma^0MU^{\gamma_5}+\gamma^0m 
	\;, \ee
and the $E_{n_0}$ of the free Hamiltonian which follows from (\ref{Hamiltonian})
by replacing $U^{\gamma_5}\to1$.  The spectrum of the Hamiltonian (\ref{Hamiltonian})
consists of an upper and a lower Dirac continuum, and --- for a strong enough pion 
field --- of a discrete bound state level of energy $E_{\rm lev}$.
By occupying the discrete level and the states of the lower continuum each 
by $N_c$ quarks in an anti-symmetric colour state and subtracting the vacuum, 
one obtains a state with unity baryon number. 
The quantity $E_{\rm sol}[U]$ is logarithmically divergent and has to be regularized 
as indicated in (\ref{soliton-energy}).
The minimization of $E_{\rm sol}[U]$ is performed for symmetry reasons 
in the so-called hedgehog Ansatz $\pi^a({\bf x})= e_r^a\;P(r)$ and determines the 
self-consistent soliton profile $P_c(r)$ and soliton field $U_c$. The nucleon mass 
$\Mn$ is given by $E_{\rm sol}[U_c]$.

Nucleon states with a definite momentum and quantum numbers are 
obtained by considering translational and rotational zero modes of the soliton.
In order to include corrections in the $1/N_c$-expansion one considers time-dependent
pion field fluctuations around the self-consistent solution. Hereby one restricts 
oneself to time-dependent rotations $U_c({\bf x})\to R(t)U_c({\bf x})R^\dag(t)$, 
where the collective coordinate $R(t)$ is a rotation matrix in SU(2)-flavour space. 
In this approximation the integral over all pion field fluctuations in 
(\ref{eff-theory}) is given by the path integral over the collective 
coordinates and solved to leading order in the collective angular velocity 
$\Omega\equiv -iR^\dag \partial_t R$. The expansion in $\Omega$ is justified, 
since the corresponding soliton moment of inertia
\be\label{Eq:mom-inertia}
    I = \frac{N_c}{6}\,\doublesum{n,\,\rm occ}{j,\,\rm non}
    \frac{\la n|\tau^a|j\ra\la j|\tau^a|n\ra}{E_j-E_n} \ee
is of ${\cal O}(N_c)$ and thus large, such that the soliton rotates slowly.
The sums in (\ref{Eq:mom-inertia}) go over occupied "occ" states~$n$
(non-occupied "non" states $j$), i.e.\  over states with $E_n\le E_{\rm lev}$
($E_j  > E_{\rm lev}$).
The $\chi$QSM  provides a particular realization of the general large-$N_c$ 
picture of the nucleon \cite{Witten:1979kh}.

With $M=350\,{\rm MeV}$ fixed from instanton phenomenology \cite{Diakonov:1983hh}
and the precise value of the cutoff (\ref{scale}) adjusted within the chosen 
regularization scheme to reproduce the experimental value of the pion decay constant, 
there are no free parameters in the $\chi$QSM. In this sense the model allows 
to evaluate in a parameter-free way nucleon matrix elements of QCD quark bilinear 
operators $\la N|\bar{\psi}(z_1)\Gamma\psi(z_2)|N\ra$ where $\Gamma$ is some 
Dirac- and flavour-matrix. In this way numerous baryon properties such as 
electromagnetic, axial or scalar form-factors, etc.,  were computed in the model
and found to agree with data to within an accuracy of typically (10-30)$\%$
\cite{Christov:1995vm,Diakonov:1988mg,Schuren:1991sc,Kubota:1999hx,Schweitzer:2003sb,Wakamatsu:2007uc,Goeke:2007fp}.
(Notice that in some studies $M$ was varied, and e.g.\ for 
$M=420\,{\rm MeV}$ in the proper-time regularization some static properties 
were found to be better described \cite{Christov:1995vm}. However, the dependence 
of the results on variations of $M$ is moderate, and within model accuracy.) 
An interesting recent application includes the extension of the model to the
description of the implicit dependence of the nucleon mass on the pion mass 
in the regime studied in lattice QCD \cite{Goeke:2005fs,Goeke:2007fq}.

The model was also applied to studies of usual 
\cite{Diakonov:1996sr,Diakonov:1997vc,Weiss:1997rt,Pobylitsa:1998tk,Wakamatsu:1998rx,Goeke:2000wv,Schweitzer:2001sr}
and generalized \cite{Petrov:1998kf,Schweitzer:2002nm,Ossmann:2004bp,Wakamatsu:2005vk}
twist-2 quark- and antiquark-distribution functions at a low scale of 
${\cal O}(\rho_{\rm av}^{-1})$. 
The small parameter in Eq.~(\ref{Eq:packing-fraction}) is of crucial importance for 
justifying the computation of twist-2 (generalized) parton distribution functions in 
the $\chi$QSM \cite{Diakonov:1995qy}. To leading order of the instanton packing 
fraction (\ref{Eq:packing-fraction}) the model quarks can be identified with QCD quark
degrees of freedom, while gluon degrees of freedom appear suppressed by this parameter 
\cite{Diakonov:1995qy}. This is essential to guarantee a consistent description, 
and the model results satisfy all general QCD requirements such as sum rules, 
inequalities, polynomiality \cite{Diakonov:1996sr,Diakonov:1997vc,Weiss:1997rt,Pobylitsa:1998tk,Wakamatsu:1998rx,Goeke:2000wv,Schweitzer:2001sr,Petrov:1998kf,Schweitzer:2002nm,Ossmann:2004bp,Wakamatsu:2005vk} 
and agree --- as far as those quantities are known --- to within (10-30)$\%$ 
with parameterizations \cite{GRV,GRSV}.

Is the model also applicable to studies of twist-3 distribution functions?
The answer to this question cannot be found within the model itself, since
twist-3 quantities can be rewritten by means of QCD equations of motion in terms 
of contributions in which gluon fields appear explicitly.
Instead it is necessary to consider this question in the instanton vacuum model,
i.e.\ in the theory from the model is derived.

The twist-3 distribution functions $g_T^a(x)$ and $h_L^a(x)$ were studied in the 
instanton vacuum model \cite{Balla:1997hf} and it was found that the pure twist-3 
parts in the Wandzura-Wilczek(-like) decompositions of these functions are strongly 
suppressed by powers of the instanton packing fraction (\ref{Eq:packing-fraction}).
As $g_T^a(x)$ and $h_L^a(x)$ can be defined without the explicit appearance of gluon 
fields --- just as $e^a(x)$ in Eq.~(\ref{Eq:def-e}) --- these functions can in 
principle be computed in models without gluon degrees of freedom \cite{Jaffe:1991ra}. 
In the $\chi$QSM this was done in \cite{Wakamatsu:2000ex} and the pure twist-3 parts 
of $g_T^a(x)$ and $h_L^a(x)$ were found to be small. Thus, in these cases the 
$\chi$QSM respects the results of the instanton vacuum model, i.e.\  of the theory 
from which it was derived. 

It would be interesting to know whether calculations of $e^a(x)$ in the $\chi$QSM are 
similarly in agreement with the instanton vacuum model. Answering this questions 
is, however, beyond the scope of this work. Here we shall take the practical
point of view of Ref.~\cite{Jaffe:1991ra}, and explore the special property
of $e^a(x)$ that allows to define it in terms of quark fields only,
without resorting explicitly to gluon degrees of freedom, which makes it 
possible to study this quantity in effective approaches with
quark and antiquark degrees of freedom only.

In the large-$N_c$ limit different flavour combinations of nucleon quantities
appear usually at different orders in the large-$N_c$ counting. 
This is also the case in the $\chi$QSM, which respects all large-$N_c$ counting rules. 
For $e^a(x)$ the isoscalar flavour combination is leading, and the isovector one 
appears only at subleading order in the $1/N_c$ expansion.
The model expressions for the different flavour combinations of $e^a(x)$ 
in proton read \cite{Schweitzer:2003uy}
\ba
	(e^u+e^d)(x) 
	&=&  	\phantom{-\,}N_c \Mn \sum\limits_{n\,\rm occ} 
		\la n|\gamma^0\delta(x\Mn-\hat{p}^3-E_n) |n\ra_{\rm reg}
		\label{Eq:model-eu+ed-occ} \\
	&=&	-\,N_c \Mn \sum\limits_{n\,\rm non} 
		\la n|\gamma^0\delta(x\Mn-\hat{p}^3-E_n) |n\ra_{\rm reg}
		\label{Eq:model-eu+ed-non} \\
	(e^u-e^d)(x) 
	&=& 	\phantom{-\,}\frac{N_c\Mn}{12I}
		\!\!\doublesum{n\,\rm occ}{j\,{\rm all},j\neq n}\!\!
		\biggl(\frac{2}{E_j-E_n}-\frac{\partial\;\;}{\partial\,x\Mn}\biggr)
		\;\la n|\tau^a|j\ra \la j|\tau^a\gamma^0
		\delta(x\Mn-\hat{p}^3-E_n) |n\ra\label{Eq:model-eu-ed-occ} \\
	&=&	-\,\frac{N_c\Mn}{12I}
		\!\!\doublesum{n\,\rm non}{j\,{\rm all},j\neq n}\!\!
		\biggl(\frac{2}{E_j-E_n}-\frac{\partial\;\;}{\partial\,x\Mn}\biggr)
		\;\la n|\tau^a|j\ra \la j|\tau^a\gamma^0
		\delta(x\Mn-\hat{p}^3-E_n) |n\ra\label{Eq:model-eu-ed-non} 
\ea
where vacuum subtractions analog to (\ref{soliton-energy}) are understood.

The possibility of computing model expressions for parton distributions 
in the two different ways, by summing over occupied 
(\ref{Eq:model-eu+ed-occ},~\ref{Eq:model-eu-ed-occ}) and non-occupied 
(\ref{Eq:model-eu+ed-non},~\ref{Eq:model-eu-ed-non}) states
is deeply connected to the analyticity properties of model expressions 
and founded on the locality of the model \cite{Diakonov:1996sr}. 

Therefore 
it is of importance to demonstrate explicitly that the equivalent formulae, 
(\ref{Eq:model-eu+ed-occ},~\ref{Eq:model-eu+ed-non}) and
(\ref{Eq:model-eu-ed-occ},~\ref{Eq:model-eu-ed-non}), 
yield respectively the same results. 
This not only provides a very useful test of the numerics.
The explicit demonstration of the ``equivalence'' of the different 
representations provides a crucial test for the internal, theoretical
consistency of the model itself, and we shall devote much effort
to this point. 

\newpage
%====== SECTION IV: (eu+ed)(x) =======================================
\section{\boldmath Calculation of $(e^u+e^d)(x)$}
\label{Sec-4:(eu+ed)}

%------ BEGIN FIGURE 1: LEVEL CONTRIBUTION to (eu+ed)(x) -----------------
\begin{wrapfigure}[16]{RD}{6.2cm}
        \centering
   	\vspace{-0.5cm}
	\includegraphics[width=5.4cm]{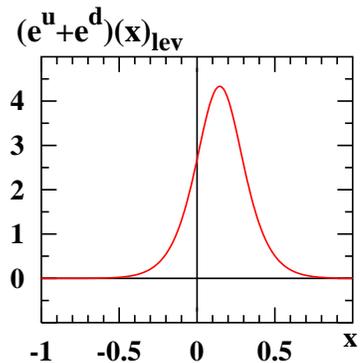}
   	\caption{\label{fig01:eu+ed-lev}
	The contribution of the discrete level 
	to $(e^u+e^d)(x)$ as function of $x$.} 
\end{wrapfigure}
%------ END FIGURE 1 -----------------------------------------------------

The isoscalar combination $(e^u+e^d)(x)$ contains a $\delta(x)$-singularity, as 
proven in \cite{Schweitzer:2003uy} and independently shown in \cite{Wakamatsu:2003uu}.
A practical procedure to cope numerically with such a singularity in the $\chi$QSM 
was suggested in \cite{Schweitzer:2003uy}, and used in \cite{Ohnishi:2003mf} 
to confirm numerically the existence of the $\delta(x)$-singularity.
In this Section we present an independent study, which will confirm the findings 
of Ref.~\cite{Ohnishi:2003mf}. Hereby we shall focus on the demonstration that 
the involved numerical calculation is well under control.

The contribution of the discrete level to $(e^u+e^d)(x)$ as well as to any quantity 
in the model UV-finite. It can be computed by directly solving the eigenvalue problem 
(\ref{Hamiltonian}) for the discrete level state \cite{Diakonov:1996sr}, or by using 
the method described below. The result is shown in Fig.~\ref{fig01:eu+ed-lev}. 

The continuum contribution can be computed in two ways, which follow from
Eqs.~(\ref{Eq:model-eu+ed-occ},~\ref{Eq:model-eu+ed-non}) (and to which we shall,
for simplicity, continue referring as sums over occupied and non-occupied states),
namely
\ba\label{Eq:model-eu+ed-cont-occ}
	(e^u+e^d)(x)_{\rm cont, \, reg} 
	=   N_c\Mn\sum\limits_{E_n<0}
	\la n|\gamma^0\delta(x\Mn-\hat{p}^3-E_n)|n\ra_{\rm reg} &,& \;\;\\
   \label{Eq:model-eu+ed-cont-non}
	=-\,N_c\Mn\sum\limits_{E_n>0}
	\la n|\gamma^0\delta(x\Mn-\hat{p}^3-E_n)|n\ra_{\rm reg} &.& \;\;\ea
It is quadratically UV-divergent and has to be regularized, as indicated in
(\ref{Eq:model-eu+ed-occ},~\ref{Eq:model-eu+ed-non}) and
(\ref{Eq:model-eu+ed-cont-occ},~\ref{Eq:model-eu+ed-cont-non}). The Pauli-Villars 
subtraction method is the only regularization known in the $\chi$QSM which
preserves all general properties of distribution functions (QCD sum rules, 
positivity, etc.) \cite{Diakonov:1996sr}.

In order to regularize the continuum contribution and to ensure the 
equivalence of the summations over occupied and non-occupied states in 
Eqs.~(\ref{Eq:model-eu+ed-cont-occ},~\ref{Eq:model-eu+ed-cont-non}), 
see \cite{Schweitzer:2003uy} for a detailed discussion, 
two Pauli-Villars subtractions are necessary 
\be\label{Eq:PV-subtractionI}
	(e^u+e^d)(x)_{\rm cont, \, reg}  
	= (e^u+e^d)(x,M)_{\rm cont}  
	- \alpha_1 (e^u+e^d)(x,M_1)_{\rm cont} 
	- \alpha_2 (e^u+e^d)(x,M_2)_{\rm cont} \ee
with 
\be\label{Eq:PV-subtractionII}
	\alpha_1 =  \frac{M}{M_1}\;\frac{M_2^2-M^2}{M_2^2-M_1^2} \;\;, \;\;\;
	\alpha_2 = -\frac{M}{M_2}\;\frac{M_1^2-M^2}{M_2^2-M_1^2} \;\;, \;\;\; 
	M_2 > M_1 > M\;.\ee
The values for the $M_i$ are fixed by regularizing the model expressions for the 
pion decay constant $f_\pi$ and the vacuum quark condensate $\la\bar\psi\psi\ra$ 
which are given in the effective theory (\ref{eff-theory}) by the (Euclidean) 
loop integrals
\ba\label{Eq:PV-fix-M1-M2}
	f_\pi^2=\int\!\frac{\di^4\pE}{(2\pi)^4}\,\frac{4N_cM^2}{(\pE^2+M^2)^2}
	\biggl|_{\rm reg} \;,\;\;\;
	\la\bar\psi\psi\ra\equiv
	\la{\rm vac}|(\bar\psi_u\psi_u+\bar\psi_d\psi_d)|{\rm vac}\ra
	= -\int\!\frac{\di^4 \pE}{(2\pi)^4}\,\frac{8N_cM}{\pE^2+M^2}
	\biggl|_{\rm reg} \; . \;\;\;\ea
The model expression for the pion decay constant is regularized and its 
phenomenological value $f_\pi=93\,{\rm MeV}$ is reproduced by a single
Pauli-Villars subtraction with $M_1=556\,{\rm MeV}$. 
Two subtractions analog to (\ref{Eq:PV-subtractionI},~\ref{Eq:PV-subtractionII}) 
are needed to regularize the model expression for the quark vacuum condensate 
(\ref{Eq:PV-fix-M1-M2}). 
In order to reproduce the phenomenological value 
$\la\bar\psi\psi\ra = -(280\pm 30)^3\,{\rm MeV}^3$ quoted in 
\cite{Gasser:1982ap} one should use $M_2=(2.1^{+1.1}_{-0.7})\,{\rm GeV}$.
For reasons which we will explain below a small value of $M_2$ is preferable, and 
we choose $M_2=986{\rm MeV}$ which yields $\la\bar\psi\psi\ra=-(220\,{\rm MeV})^3$. 
This is sufficiently close to the phenomenological value --- considering the 
typical accuracy of the model.

In order to evaluate $(e^u+e^d)(x)$ we use the following procedure
\cite{Diakonov:1997vc}. 
For the numerical calculation the Hamiltonian (\ref{Hamiltonian}) is expressed and 
diagonalized in the basis of the eigenstates of the free Hamiltonian. The spectrum of 
(\ref{Hamiltonian}) is discretized by placing the soliton in a finite but sufficiently
large spherical box of the size $R_{\rm box}$ and imposing appropriate boundary 
conditions \cite{Kahana:1984be}. 
The spectrum is made finite by cutting off quark momenta above some large numerical 
cutoff $\Lambda_{\rm num}$ chosen much larger than any other (physical or numerical) 
scale involved in the problem. We use $R_{\rm box}=(8-12)\,{\rm fm}$ and 
$\Lambda_{\rm num}=(8-9)\,{\rm GeV}$.

To compute the continuum contribution we introduce an intermediate regularization 
for any of the contributions in Eq.~(\ref{Eq:PV-subtractionI}) schematically as
\be\label{Eq:intermediate-reg}
	(e^u+e^d)(x,M_i) = N_c \Mn \biggl[\sum\limits_n 
	\la n|\gamma^0\delta(x\Mn-\hat{p}^3-E_n)|n\ra\;R(E_n,\,\Lambda)\biggr]_{M_i}
	\;\ee
where it is understood that the expression in squared brackets is to be evaluated 
with the Hamiltonian (\ref{Hamiltonian}) if $M_i=M$, or with Hamiltonians analog to 
(\ref{Hamiltonian}) but with $M$ replaced by $M_1$ or $M_2$, and where vacuum 
subtraction is implied. In (\ref{Eq:intermediate-reg}) $R(\omega,\,\Lambda)$ is a 
smooth regulator function with $R(0,\,\Lambda)=1$ and $R(\omega,\,\Lambda)\to0$ for 
$|\omega|\to\infty$. The 
intermediate cutoff $\Lambda$ must satisfy $M_i \ll \Lambda \ll \Lambda_{\rm num}$.
It is due to this condition that we prefer a low value of $M_2$, see above.
In practice we use $\Lambda\sim(3-6)\;{\rm GeV}$.
The regulator function can be chosen to be of e.g.\ Gaussian or Wood-Saxon type. 
The dependence on $\Lambda$ and the choice of $R$ is removed at the end of the 
calculation by an extrapolation procedure.

It is convenient to turn (\ref{Eq:intermediate-reg}) into a spherically symmetric
form by replacing $\hat{p}^3\to{\bf u}\cdot\hat{\bf p}$ where ${\bf u}$ is a unit 
vector. (We recall that the 3-direction was singled out arbitrarily by choosing the 
spatial component of the light-like vector $n$ in (\ref{Eq:def-e}) along that axis.) 
Averaging over the possible orientations of ${\bf u}$ yields
\be\label{Eq:sperical-symm}
	(e^u+e^d)(x,M_i)
	=N_c\Mn\biggl[\sum\limits_n\la n|\gamma^0\;\frac{1}{2|\hat{\bf p}|}\,
	  \Theta(|\hat{\bf p}|-|x\Mn-E_n|)|n\ra\;
	  R(E_n,\,\Lambda)\biggr]_{M_i}\;.\ee
As we work in a discrete basis the expression (\ref{Eq:sperical-symm}) is a 
discontinuous function of $x$ due to the appearance of the $\Theta$-function,
and would become continuous only in the infinite volume limit. Rather than trying
to take this limit numerically, which would be a time-consuming procedure,
one may instead smear the expression in (\ref{Eq:sperical-symm}) by 
convoluting it with a narrow Gaussian as 
\be\label{Eq:smearing}
	(e^u+e^d)(x,M_i) = 
	\frac{N_c\Mn}{\gamma\pi^{1/2}}\int\di x^\prime\,e^{-(x-x^\prime)^2\gamma^{-2}}
	\;\biggl[\sum\limits_n\la n|\gamma^0\;\frac{1}{2|\hat{\bf p}|}\,
	  \Theta(|\hat{\bf p}|-|x^\prime\Mn-E_n|)|n\ra\;
	  R(E_n,\,\Lambda)\biggr]_{M_i}\;.\ee
In the limit $\gamma\to 0$ one recovers the original expression 
(\ref{Eq:sperical-symm}). The parameter $\gamma>0$ has to be chosen such that 
it is, on the one hand, sufficiently large compared to the typical splitting 
of energy levels in the discretized spectrum,
% which is proportional to $R_{\rm box}^{-1}$, 
and on the other hand, sufficiently small such that the ``smeared function'' 
is still a good approximation to the true result.
For our box sizes $\gamma = 0.1$ is adequate \cite{Diakonov:1997vc}. 
In the end of the day the smearing can be removed by a deconvolution 
procedure, though in practice one finds that continuous functions are 
sufficiently well approximated by (\ref{Eq:smearing}).

%------ BEGIN FIGURE 2: (eu+ed)(x) --------------------------------------
%
\begin{figure*}[b!]
\begin{tabular}{ccc}
\includegraphics[width=5.4cm]{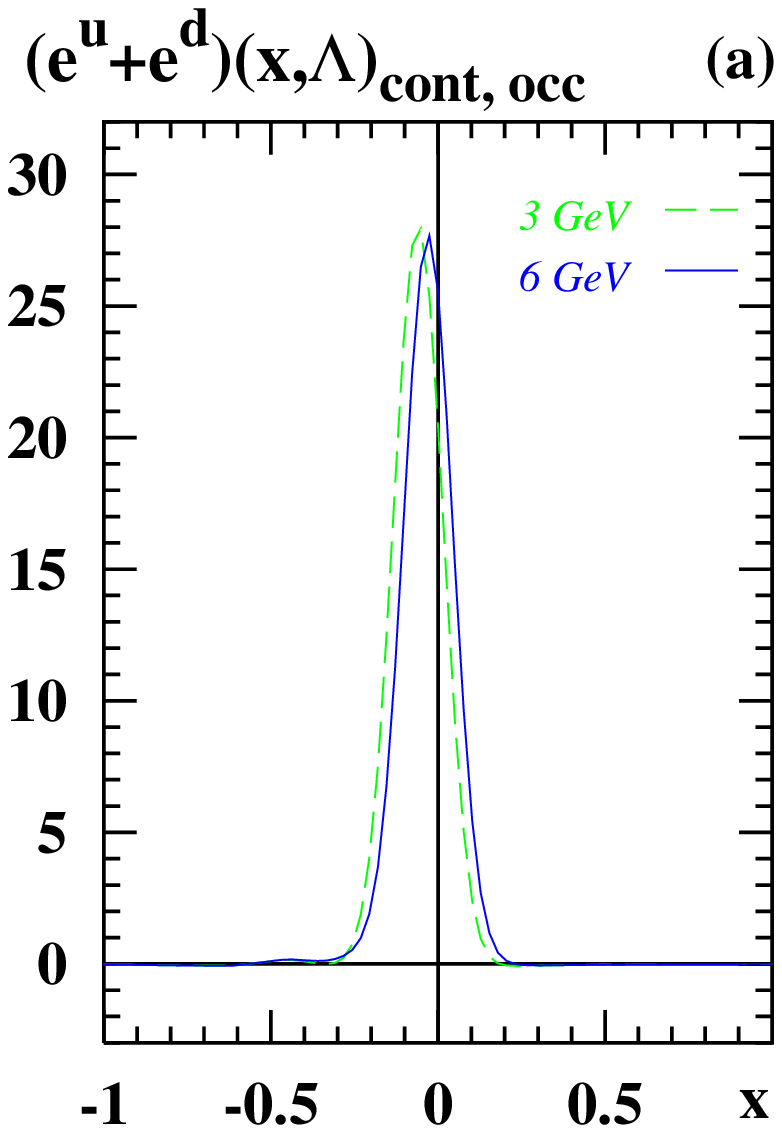} &
\includegraphics[width=5.4cm]{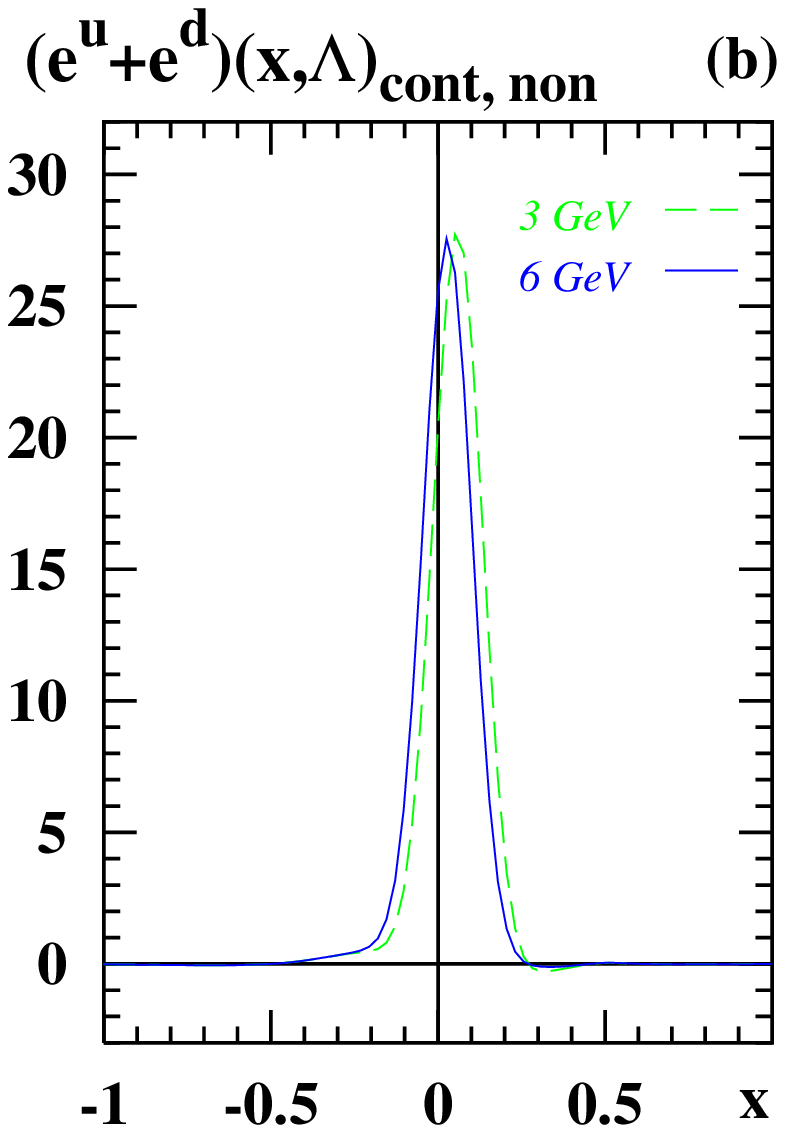} &
\includegraphics[width=5.4cm]{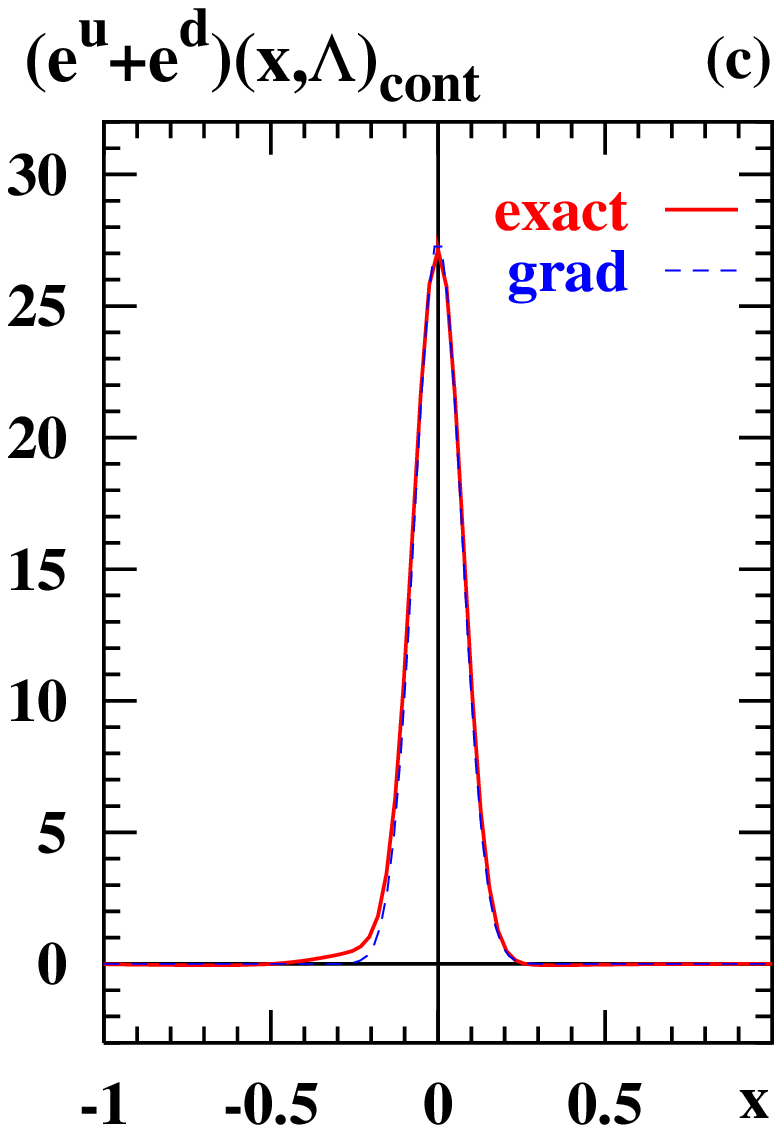}
\end{tabular}
	\caption{\label{fig02:eu+ed}
	The continuum contribution to $(e^u+e^d)(x)$ as function of $x$. 
	(a) The result from the sum over occupied levels 
	(\ref{Eq:model-eu+ed-cont-occ}) for different values 
	of the intermediate cutoff $\Lambda$. 
	(c) The same as (b) but from the sum over non-occupied levels
	(\ref{Eq:model-eu+ed-cont-non}).
	(d) Solid line: the final result obtained after the extrapolation 
	$\Lambda\to\infty$. Dotted line: the result from 
	the gradient expansion \cite{Schweitzer:2003uy}, see text.}
\end{figure*}
%
%------ END FIGURE 2 -----------------------------------------------------

It is precisely the smearing procedure which enables one to cope numerically with 
a $\delta$-function-type singularity. In fact, the smearing trick, turns the 
$\delta(x)$-contribution in $(e^u+e^d)(x)$ into a narrow-Gausssian of a well-defined 
width $\gamma$ centered around $x=0$.
Using the self-consistent profile from \cite{Weiss:1997rt} which yields 
$1140\,{\rm MeV}$ for the soliton energy and the above described parameters we obtain 
for the regularized continuum contribution $(e^u+e^d)(x,\Lambda)_{\rm cont}$ for the 
intermediate cutoffs $\Lambda=3$ and $6\,{\rm GeV}$ the results shown in 
Figs.~\ref{fig02:eu+ed}a and b.

The results depend on whether one sums over occupied,
Eq.~(\ref{Eq:model-eu+ed-cont-occ}), or non-occupied, 
Eq.~(\ref{Eq:model-eu+ed-cont-non}), states and on the intermediate cutoff $\Lambda$. 
However, after extrapolating  $\Lambda\to\infty$ both formulae 
(\ref{Eq:model-eu+ed-cont-occ}) and (\ref{Eq:model-eu+ed-cont-non})
yield within a numerical accuracy of $1\%$ the same result, 
which is shown in Fig.~\ref{fig02:eu+ed}c as solid line.

%====== SECTION V: (eu+ed)(x) and delta(x) ===========================
\section{\boldmath The origin of the $\delta(x)$-singularity in $(e^u+e^d)(x)$}
\label{Sec-5:(eu+ed)-and-delta}

Having convinced ourselves that one obtains for the continuum contribution 
of $(e^u+e^d)(x)$ the same result, irrespective whether one computes it
by means of (\ref{Eq:model-eu+ed-cont-occ}) or (\ref{Eq:model-eu+ed-cont-non}),
we have confidence in our numerical results, and are in the position to
address the question: What precisely gives rise to the $\delta(x)$-singularity 
in $(e^u+e^d)(x)$ in the $\chi$QSM?

As there is apparently no $\delta(x)$-term in the level part, 
see Fig.~\ref{fig01:eu+ed-lev}, one has to focus on the continuum contribution.
The continuum contribution 
(\ref{Eq:model-eu+ed-cont-occ}) or (\ref{Eq:model-eu+ed-cont-non}), is given 
in the gradient expansion by \cite{Schweitzer:2003uy} 
\be\label{Eq:eu+ed-cont-grad}
	(e^u+e^d)(x)_{\rm cont} = C\;\delta(x) + {\cal O}(\nabla U)
	\;,\;\;\; 
	C= %\la\bar\psi\psi\ra \int\!\di^3{\bf r}\;\frac12\,{\rm tr}_F
	   %\biggl(\frac{U+U^\dag\!}{2}-1\biggr) =
	   \la\bar\psi\psi\ra \int\!\di^3{\bf r}\;\biggl(\cos P(r)-1\biggr) \;,
\ee
where ${\cal O}(\nabla U)$ denotes terms which 
{\sl (i)} contain one or more gradients of the chiral field $U$,
i.e.\  are suppressed in a chiral expansion, and
{\sl (ii)} are regular functions of $x$.

The result (\ref{Eq:eu+ed-cont-grad}) is remarkable for two reasons.
First, the coefficient $C$ of the $\delta(x)$-function can be computed 
{\sl exactly}. Second, the quadratic and logarithmic divergences appearing in 
$C$ can be reexpressed in terms of $\la\bar\psi\psi\ra$ \cite{Schweitzer:2003uy}. 

It is is interesting to confront (\ref{Eq:eu+ed-cont-grad}) which describes 
the singular part of $(e^u+e^d)(x)$ in the gradient expansion exactly with 
our numerical result. For that we evaluate $C$ using the same Pauli-Villars 
masses, i.e.\  $\la\bar\psi\psi\ra=-(220\,{\rm MeV})^3$,
and the same self-consistent profile $P(r)$ as in the numerical 
calculation which yields 
\be\label{Eq:C-grad-number}
	C_{\rm here} = 4.88\;.
\ee
(The introduced index ``here'' reminds that this result 
holds for the Pauli-Villars masses used here, in this calculation.)
Moreover, we smear the $\delta(x)$-function according 
to (\ref{Eq:smearing}) with the same parameter $\gamma=0.1$ as used in the numerics.
In this way we obtain the result shown as dashed line in Fig.~\ref{fig02:eu+ed}c.

The impressive agreement in Fig.~\ref{fig02:eu+ed}c fully confirms the findings of 
Ref.~\cite{Schweitzer:2003uy} that the $\delta(x)$-contribution in the $\chi$QSM 
originates {\sl solely} from the leading order of the gradient expansion where 
it is related to the quark vacuum condensate. 
Fig.~\ref{fig02:eu+ed}c also shows that the regular contribution to the 
continuum part of $(e^u+e^d)(x)$ is small (on the scale in Fig.~\ref{fig02:eu+ed}c).
We observed similar agreements by varying the numerical parameters 
(Pauli-Villars masses $M_i$, profile $P(r)$, $\gamma$, etc.). 
These observations illustrate the utility of the gradient 
expansion as a powerful tool to control the numerics. 

Let us report the following detail which further increases our faith into the 
quality of the numerics. In order to achieve the satisfactory agreement in 
Fig.~\ref{fig02:eu+ed}c we considered the finite size of the spherical box
used in the numerics, and integrated in (\ref{Eq:eu+ed-cont-grad}) over the 
radial component $r=|{\bf r}|$ only up to $r=R_{\rm box}=12\,{\rm fm}$
which gave the result in (\ref{Eq:C-grad-number}). 
In the chiral limit the profile function behaves as $P(r)=A/r^2$ at large $r$,
where $A$ is related to the isovector axial coupling constant $g_A$ by 
$A = (3 g_A)/(8 \pi f_\pi^2)$. In practice the asymptotics sets in already 
for $r\gtrsim 3\,{\rm fm}$ \cite{Weiss:1997rt}. Thus, the finite size effect 
for the coefficient $C$ in (\ref{Eq:eu+ed-cont-grad}) is
\be\label{Eq:finite-size-effect}
	\delta C
	= \la\bar\psi\psi\ra \;
	4\pi\int_{R_{\rm box}}^\infty\!\!\di r\;r^2\biggl(\cos P(r)-1\biggr)
	= -\,\la\bar\psi\psi\ra \;\frac{2\pi\,A^2}{R_{\rm box}}\;.
\ee
The finite size effect slowly vanishes with increasing box size.
For $R_{\rm box}=12\,{\rm fm}$ one has $\delta C_{\rm here}= 0.196$, i.e.\  
the result for $C_{\rm here}$ in (\ref{Eq:C-grad-number}) is about $4\%$ smaller 
compared to its infinite volume limit. Neglecting this effect would 
yield a clearly visible mismatch in Fig.~\ref{fig02:eu+ed}c.
Thus, as a byproduct we see that finite box size effects in our calculation 
are of the order of magnitude of a few percent, which is acceptable considering the
typical model accuracy.

Thus, besides confirming numerically the presence of a $\delta(x)$-contribution 
in $(e^u+e^d)(x)$ as was done previously in \cite{Ohnishi:2003mf},
we furthermore have numerically confirmed the fact that the coefficient $C$ of 
the $\delta(x)$-function can be computed {\sl exactly} using gradient expansion 
\cite{Schweitzer:2003uy}. This is of importance because 
only the gradient expansion allows to relate $C$ to the quark vacuum condensate.

The exact prediction for the coefficient $C$ in the $\chi$QSM is therefore
\be\label{Eq:C-exact}
	C_{\rm exact} = 10.5^{+3.7}_{-3.1}
\ee
using the self-consistent profile \cite{Weiss:1997rt} and the phenomenological 
value $\la\bar\psi\psi\ra = -(280\pm 30)^3\,{\rm MeV}^3$ \cite{Gasser:1982ap}
whose uncertainty yields to the error shown in (\ref{Eq:C-exact}).

We remark that in \cite{Ohnishi:2003mf}, where the Pauli-Villars masses were
fixed to reproduce $\la\bar\psi\psi\ra=-(280\,{\rm MeV})^3$, the result $C\simeq 9.9$ 
was obtained numerically. This is in good agreement with the central value of
(\ref{Eq:C-exact}) and the small discrepancy, apart from numerical uncertainties
(finite box size effects),
is due to the slightly different value of the constituent mass $M=375\,{\rm MeV}$ 
used in \cite{Ohnishi:2003mf}, which yields a somehow different self-consistent 
profile.
We checked that using larger values of $M_2\sim 1.5\,{\rm GeV}$ we are able to 
reproduce larger values of $C$. The price to pay for that is, however, a worse 
numerical stability because then the required hierarchy 
$M_i\ll \Lambda\ll \Lambda_{\rm num}$ holds less satisfactorily,
unless one increases $\Lambda$ and $\Lambda_{\rm num}$ accordingly
from which we refrained being limited by the computing  resources available to us.

Having established the presence of a $\delta(x)$-type-singularity in $(e^u+e^d)(x)$,
confirming thereby the findings of \cite{Ohnishi:2003mf}, we now turn to the task
of computing the regular part of the continuum contribution. 

\newpage
%====== SECTION VI: x*(eu+ed)(x) =====================================
\section{\boldmath Calculation of $x(e^u+e^d)(x)$}
\label{Sec-6:x(eu+ed)}

Being interested in the regular part of the continuum contribution to 
$(e^u+e^d)(x)$ the method of Sec.~\ref{Sec-4:(eu+ed)} is not adequate.
The smearing trick, which allows to visualize the delta-function, is of disadvantage
in this case. It turns the $\delta(x)$-function into a Gaussian that
penetrates into the regions $x\neq 0$ and dominates there over the regular part, 
Fig.~\ref{fig02:eu+ed}c. 
Thus, a reliable computation of the regular part requires a different technic. 
Here we shall compute $x(e^u+e^d)(x)$ where the $\delta(x)$ drops out.
(Due to the smearing procedure (\ref{Eq:smearing}) computing $x(e^u+e^d)(x)$ is, of 
course, not the same as computing $(e^u+e^d)(x)$ and multiplying the result by $x$.)

First we have to clarify how to regularize $x(e^u+e^d)(x)$. 
It is worth to reconsider this point because $x(e^u+e^d)(x)$ and $(e^u+e^d)(x)$ 
differ by the appearance of a $\delta(x)$-term. As shown in \cite{Schweitzer:2003uy}, 
the double subtraction (\ref{Eq:PV-subtractionI},~\ref{Eq:PV-subtractionII}) is 
needed to make finite the {\sl coefficient} of the $\delta(x)$-function. 
But is it also necessary to regularize $x(e^u+e^d)(x)$? It is worth to reconsider
this point: one could save a lot of computing time, if no or only one subtraction 
were necessary.

By going step by step through the Eqs.~(32-40) of Ref.~\cite{Schweitzer:2003uy}, 
one believes, at a first impression, that a single subtraction 
(corresponding to (\ref{Eq:PV-subtractionI},~\ref{Eq:PV-subtractionII}) 
for $M_2\to\infty$) is {\sl enough} to remove from  $x(e^u+e^d)(x)$
a quadratic divergence and restore the equivalence of
summations over occupied and non-occupied levels \cite{Schweitzer:2003uy}. 
However, such a  conclusion is  premature and we show here that a 
double subtraction is adequate.  

In order to illustrate why we integrate $x(e^u+e^d)(x)$ from 
(\ref{Eq:model-eu+ed-cont-occ},~\ref{Eq:model-eu+ed-cont-non}) over $x$, explore the 
model equations of motion, and arrive at expressions for the continuum contribution 
to the second moment of $(e^u+e^d)(x)$ which read  \cite{Schweitzer:2003uy}
\ba
\label{Eeq:2mom-grad-00-occ}
	\int\di x\,x(e^u+e^d)(x)_{\rm cont} 
	&=&   \frac{N_cM}{2\Mn} \sum_{E_n<0} \la n|(U+U^\dag)|n\ra_{\rm reg}  \\
\label{Eeq:2mom-grad-00-non}
	&=& - \frac{N_cM}{2\Mn} \sum_{E_n>0} \la n|(U+U^\dag)|n\ra_{\rm reg}  \;.\ea
Clearly, the difference of the two expressions in (\ref{Eeq:2mom-grad-00-occ}) and 
(\ref{Eeq:2mom-grad-00-non}) must be zero, i.e.\  one expects
\be\label{Eq:2mom-grad-02}
	A_{\rm reg}\equiv
	\frac{N_cM}{2\Mn}\sum_{n\,\rm all}\la n|(U+U^\dag)|n\ra_{\rm reg}
	= \frac{N_cM}{\Mn}\;{\rm Sp}\,\Biggl[\frac{U+U^\dag}{2}-1\Biggr]_{\rm reg}
	\stackrel{!}{=} 0 \;.\ee
Here ${\rm Sp}$ denotes the functional trace which can be saturated by any complete 
set of functions, ${\rm Sp}[\,\dots\,]=\sum_n\la n|\dots|n\ra$ being one example. 
Notice that the ``$-1$'' in Eq.~(\ref{Eq:2mom-grad-02}) is due to the 
explicitly included vacuum subtraction. %, and that $U+U^\dag=2\cos P(r)$.

In the numerical calculation the expression for $A_{\rm reg}$ is evaluated
with an intermediate regularization, see Eq.~(\ref{Eq:intermediate-reg}),
such that in the intermediate step we are interested in the following expression
\be\label{Eq:2mom-grad-03}
	A(\Lambda,M) = \frac{N_cM}{\Mn}\;{\rm Sp}\,
	\Biggl[\frac{U+U^\dag}{2}\,R(H,\,\Lambda)
	-R(H_0,\,\Lambda)\Biggr] \;.\ee
We evaluate $A(\Lambda,M)$ in Eq.~(\ref{Eq:2mom-grad-03}) in gradient expansion 
where $R(H,\,\Lambda)=R(H_0,\,\Lambda)+{\cal O}(\nabla U)$ holds, and saturate 
${\rm Sp}[\,\dots\,]=$
$\int\frac{\di^3{\bf k}}{(2\pi)^3}\la {\bf k}|{\rm tr}\,\dots|{\bf k}\ra$ in the 
spectrum of the free momentum operator where ${\rm tr}$ denotes the trace over 
Dirac- and flavour-indices. As positive and negative energies must 
be regularized equally, i.e.\ 
$R(\omega,\Lambda)=R(-\omega,\Lambda)$, we obtain:
\be\label{Eq:2mom-grad-04}
	A(\Lambda,M) 
	= \frac{8N_c M}{\Mn}\;
	\int\!\di^3{\bf x}\,(\cos P(r)-1)\;
	\int\!\frac{\di^3{\bf k}}{(2\pi)^3}\,	
	R(\sqrt{{\bf k}^2+M^2},\Lambda)\;.
% FOR GAUSS REGULATOR:
%	= \frac{8N_c M}{\Mn}\;
%	\int\di^3{\bf x}\biggl(\cos P(r)-1\biggr)
%	\int\frac{\di^3{\bf k}}{(2\pi)^3}\,	
%	\exp\biggl(-\frac{{\bf k}^2+M^2}{\Lambda^2}\biggr)
%	= \frac{N_c M\Lambda^3}{\pi^{3/2}\Mn}\;
%	\exp\biggl(-\frac{M^2}{\Lambda^2}\biggr)
%	\int\di^3{\bf x}\biggl(\cos P(r)-1\biggr) + {\cal O}(\nabla U)
	\ee
We recognize, most easily by employing a definite regulator such as e.g.\ 
a simple Gaussian $R(\omega,\Lambda)=\exp(-\omega^2/\Lambda^2)$, that 
$A(\Lambda,M)$ has a cubic and a linear divergence for $\Lambda\to\infty$.
The double subtraction (\ref{Eq:PV-subtractionI},~\ref{Eq:PV-subtractionII})
not only removes these divergences but also restores the equivalence of the 
summations over occupied or non-occupied states, i.e.\ 
\be\label{Eq:2mom-grad-05}
	A_{\rm reg} \equiv \lim\limits_{\Lambda\to\infty} A(\Lambda) = 0
	\;,\;\;\;\mbox{where}\;\;\;
	A(\Lambda)\equiv A(\Lambda,M) - \alpha_1 A(\Lambda,M_1)
	- \alpha_2 A(\Lambda,M_2)\;.\ee
One can convince oneself similarly as done in \cite{Schweitzer:2003uy} that 
higher orders in $\nabla U$ omitted in (\ref{Eq:2mom-grad-04}) do not spoil the 
above argumentation. 
Thus, $x(e^u+e^d)(x)$ must be regularized exactly in the same way as $(e^u+e^d)(x)$ 
according to (\ref{Eq:PV-subtractionI},~\ref{Eq:PV-subtractionII}).

The reason why the calculation in Eqs.~(32-40) of \cite{Schweitzer:2003uy} lead us to 
the premature conclusion that a single subtraction could be sufficient is because
here and in \cite{Schweitzer:2003uy} different matrix 
elements were evaluated in gradient expansion, namely
\be
	\la n| E_n\gamma^0|n\ra
	\biggl|_{\mbox{\footnotesize used in \cite{Schweitzer:2003uy}}}
	\;\;\stackrel{\rm eom}{=}\; 
	\frac12\,   \la n|\{\hat{H},\gamma^0\}|n\ra
	= \frac12\,M\,\la n|(U+U^\dag)|n\ra\biggl|_{\mbox{\footnotesize used here}}
\ee
which are connected by equations of motion (eom).
The latter are, of course, not respected in a truncated expansion.
Another example that the gradient expansion yields results at variance with
eom is documented in Sec.~7 of Ref.~\cite{Diakonov:1996sr}.

%------ BEGIN FIGURE 3: CONTINUUM OF SMEARED x*(eu+ed)(x) -----------------
\begin{wrapfigure}[36]{RD}{6.2cm}
        \centering
   	\vspace{-0.7cm}
\includegraphics[width=5.4cm]{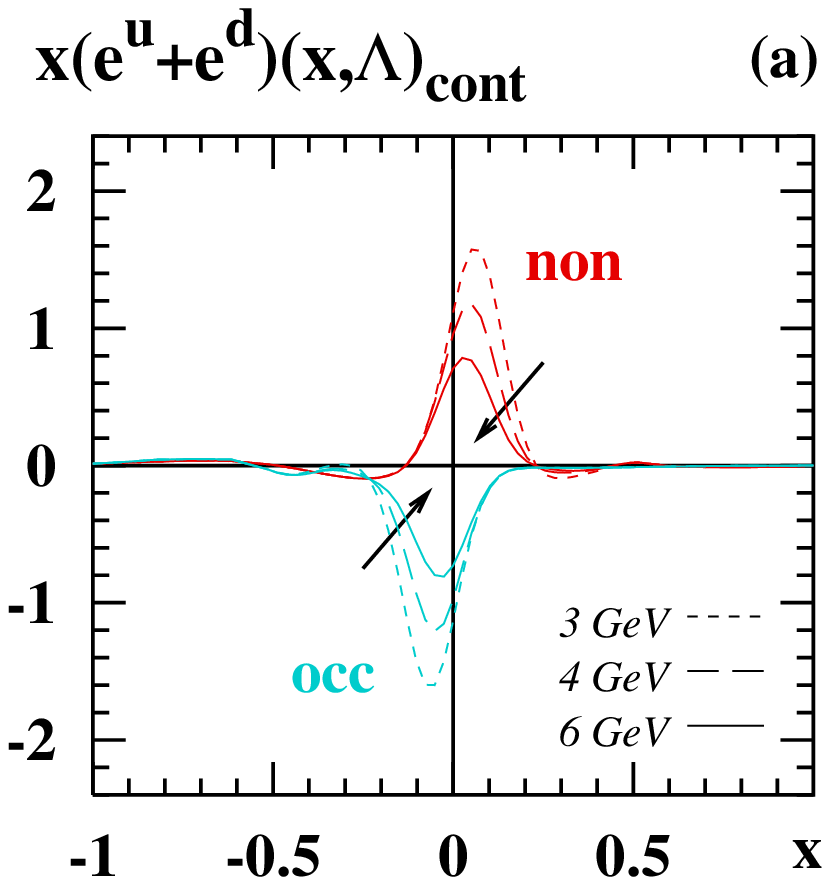}
	\caption{\label{FigS:x(eu+ed)}
	The continuum contribution to $x(e^u+e^d)(x)$ as function of $x$
	from sums over occupied (``occ'') and non-occupied (``non'')
	levels for different intermediate cutoffs $\Lambda$. 
	The arrows indicate the tendency of the curves with increasing cutoff.} 
%\end{wrapfigure}%\end{figure*}
%
%------ END FIGURE 3 -----------------------------------------------------
%
%
%------ BEGIN FIGURE 4: ANOMALY IN \int\di x x*(eu+ed)(x) -----------------
%
%\begin{wrapfigure}[18]{RD}{6.2cm}
        \centering
\includegraphics[width=5.4cm]{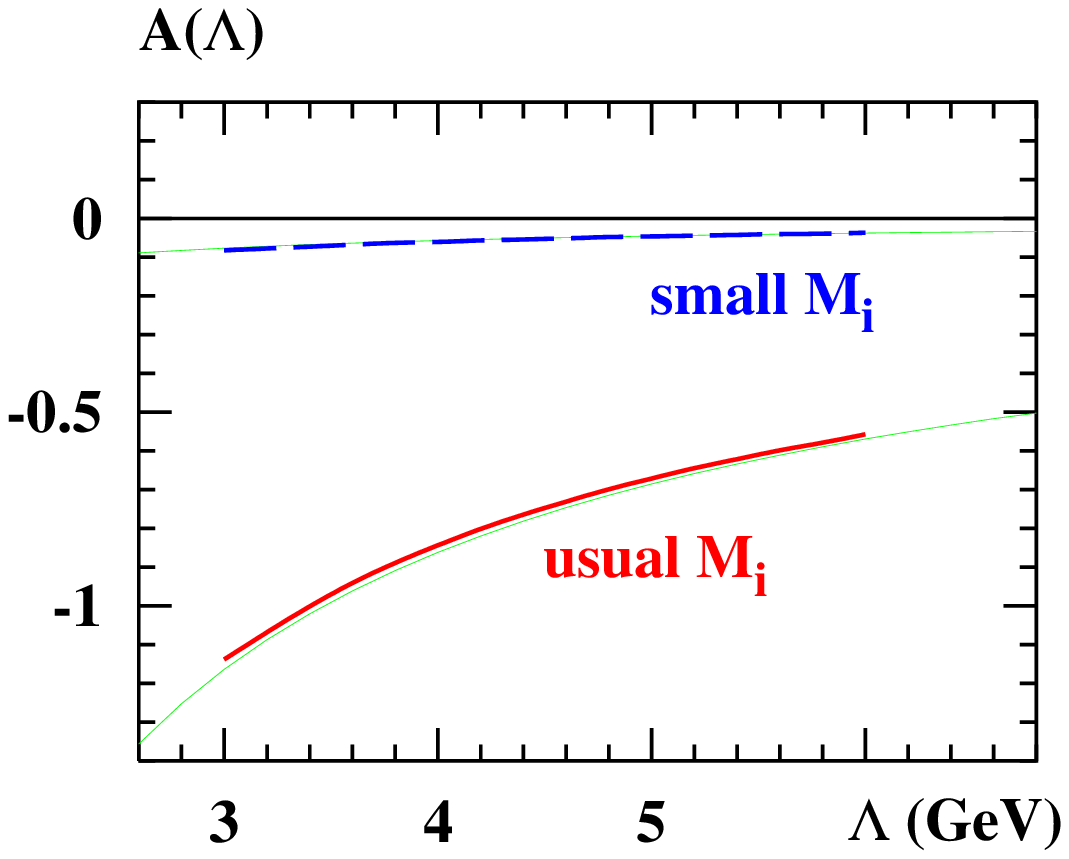}
	\caption{\label{Fig04:anomaly-2mom}
	The difference $A(\Lambda)$ defined in (\ref{Eq:2mom-grad-02})
	as function of the intermediate cutoff $\Lambda$. 
	It must vanish for $\Lambda\to\infty$ according to (\ref{Eq:2mom-grad-05}).
	The thick curves are the numerical results.
	Solid: for ``usual'' Pauli-Villars masses used throughout.
	Dashed: for a set of smaller masses $M_1=389\,{\rm MeV}$ and 
	$M_2=700\,{\rm MeV}$.
	The thin curves are the respective analytical results from 
	Eq.~(\ref{Eq:2mom-grad-04}).
	} 
\end{wrapfigure}%\end{figure*}
%
%------ END FIGURE 4 -----------------------------------------------------

Having clarified the issue of regularization we turn to the numerical evaluation.
 At first glance one may have the impression that different expressions exist, namely
\ba\label{Eq:model-x(eu+ed)}
	x(e^u+e^d)(x)  
	&=& x\Mn N_c\sum\limits_{n\;\rm occ} 
	  \la n|\gamma^0\delta(x\Mn-\hat{p}^3-E_n)|n\ra_{\rm reg} \\
	&=& N_c\sum\limits_{n\;\rm occ}
	  \la n|\gamma^0\delta(x\Mn-\hat{p}^3-E_n)\,(\hat{p}^3+E_n)\,|n\ra_{\rm reg}
	\nonumber
	\ea
and analog expressions with summations over non-occupied states. However, 
it is gratifying to observe that after averaging over directions either 
of these expressions just yields $x$ times the result in 
Eq.~(\ref{Eq:sperical-symm}). 

The numerical procedure of Sec.~\ref{Sec-4:(eu+ed)} yields for 
$x(e^u+e^d)(x,\Lambda)_{\rm cont}$ the results shown in Fig.~\ref{FigS:x(eu+ed)}. 
We make two interesting observations.
First, compared to the huge effect of the smeared out $\delta(x)$-function in
Fig.~\ref{fig02:eu+ed} the regular contribution is small.
Second, the results for the sums over positive and negative energy states
exhibit a tendency with increasing intermediate cutoff $\Lambda$ 
to converge slowly towards a common result. However, this convergence is remarkably 
slow, and the numerical stability of a $\Lambda\to\infty$ extrapolation is poor, 
especially for small $|x|\lesssim 0.2$.

The slow convergence can be understood as follows. We compute the moment 
$\int\di x\,x(e^u+e^d)(x)_{\rm cont}$ summing over respectively occupied,
Eq.~(\ref{Eeq:2mom-grad-00-occ}), and non-occupied, Eq.~(\ref{Eeq:2mom-grad-00-non}),
states, and take the difference. This yields the $A(\Lambda)$ defined in 
Eq.~(\ref{Eq:2mom-grad-05}), which is expected to be zero in the limit $\Lambda\to 0$. 
The numerical results for $A(\Lambda)$ for $3\,{\rm GeV}\le\Lambda\le 6\,{\rm GeV}$
are shown in Fig.~\ref{Fig04:anomaly-2mom}. We see that $A(\Lambda)$ tends to zero 
with increasing $\Lambda$ but very slowly. 
(We checked that one obtains within few percent the same result from integrating 
the curves for $x(e^u+e^d)(x)_{\rm cont}$ in Fig.~\ref{FigS:x(eu+ed)}.)

%
% HERE:
%
This slow vanishing of the ``anomaly'' (in the sense of \cite{Goeke:2000wv})
$A(\Lambda)$ is, in fact, precisely what we expect. In order to see that, 
we evaluate the theoretical expression (\ref{Eq:2mom-grad-04}) for $A(\Lambda)$ 
in the same way as done in the numerics, namely for the Wood-Saxon regulator
\be
        R(\omega,\Lambda) = \frac{1}{1+\exp[(|\omega|-\Lambda)/\varepsilon]}
\ee
with the box-size dependent width $\varepsilon = 4/R_{\rm box}$ 
(here $R_{\rm box}=12\,{\rm fm}$), and integrate 
in spherical coordinates over ${\bf k}$ up to 
$|{\bf k}| \le \Lambda_{\rm num} = 8.9\,{\rm GeV}$.
The result is shown in Fig.~\ref{Fig04:anomaly-2mom} as dashed line, 
and we observe again an impressive agreement of the analytical and 
numerical results.

To test further the theoretical result (\ref{Eq:2mom-grad-04})
we repeated the calculation with $M_1=389\,{\rm MeV}$ and 
$M_2=700\,{\rm MeV}$.
With smaller (compared to the ``usual'' ones fixed in the sequence of 
Eq.~(\ref{Eq:PV-fix-M1-M2}) and used throughout) Pauli-Villars masses 
one subtracts ``more''. 
(The opposite is evident, for $M_{1,2}\to\infty$ one recovers unregularized, 
divergent expressions.) Therefore, continuum contributions and consequently 
$A(\Lambda)$ are smaller. Also in this case the analytical 
and numerical results agree remarkably, see Fig.~\ref{Fig04:anomaly-2mom}.

It happens that $A(\Lambda)$ is much larger then the continuum contribution 
whose computation it hampers,  namely
(in the numerator of the undesired, anomalous terms cancel out)
\be\label{Eq:relative-anomaly}
	\left|\frac{
	 \int\di x\,x(e^u+e^d)(x,\Lambda)_{\rm cont,\,occ}
	-\int\di x\,x(e^u+e^d)(x,\Lambda)_{\rm cont,\,non}}
     	{\int\di x\,x(e^u+e^d)(x,\Lambda)_{\rm cont,\,occ}
	+\int\di x\,x(e^u+e^d)(x,\Lambda)_{\rm cont,\,non}}\right|
	\gtrsim \cases{10 & for usual Pauli-Villars masses\cr
			3 & for small Pauli-Villars masses}
\ee
for $3\,{\rm GeV}\le\Lambda\le 6\,{\rm GeV}$. The result (\ref{Eq:relative-anomaly}) 
indicates that, the better the condition $M_i \ll \Lambda \ll \Lambda_{\rm num}$
is realized, the less the undesired anomalous difference is disturbing.
Thus, were we able to use intermediate cutoffs satisfying more convincingly 
the condition $\Lambda\gg M_i$, it would be possible to perform reliably 
the extrapolation $\Lambda\to\infty$. However, here we are restricted, 
since a numerical cutoff $\Lambda_{\rm num}$ of substantially more than $10\,{\rm GeV}$
would require unacceptably long computing times.
Unfortunately, this means that with justifiable effort one cannot establish 
the equivalence of summations over occupied and non-occupied levels.

The calculation in Eqs.~(32-40) of \cite{Schweitzer:2003uy} reveals how 
this contribution, which only slowly vanishes with increasing $\Lambda$, 
is distributed in $x$. The anomalous terms appear at $x<0$ ($x>0$) when 
computing $x(e^u+e^d)(x,\Lambda,M_i)$ from occupied (non-occupied) states.
Due to the smearing procedure, however, the effects of the anomalous terms 
penetrate also in the respectively opposite $x$-regions, and yield 
the picture in Fig.~\ref{FigS:x(eu+ed)}.

>From this observation it is clear how one can evaluate $x(e^u+e^d)(x)$.
We have to switch off the smearing, and for $x>0$ ($x<0$) we have
to sum over occupied (non-occupied) states. 

The observation that for $x>0$ ($x<0$) the sums over occupied 
(non-occupied) states converge faster than in the opposite $x$-regions
was already noticed in \cite{Diakonov:1997vc}. 
Still, in all examples encountered so far, the convergence in those
slower-convergence-$x$-regions was fast enough in order to establish reliably 
the equivalence of results from summations over occupied and non-occupied states
\cite{Diakonov:1997vc,Weiss:1997rt,Pobylitsa:1998tk,Goeke:2000wv,Schweitzer:2001sr,Ossmann:2004bp}.
Here we face for the first time the situation where this is not possible,
which is not surprizing as it is also the first time one has to deal with
Pauli-Villars-masses of ${\cal O}(1\,{\rm GeV})$, and we are forced to 
give up this strong test of the numerical results.

The price to pay for giving up the smearing step (\ref{Eq:smearing}) is less serious. 
One may turn discontinuous results into final, continuous ones, see below.

Fig.~\ref{Fig5:x(eu+ed)-unsmeared-calculation}a shows the results 
for the continuum contribution to the antiquark distribution, 
$x(e^{\bar u}+e^{\bar d})(x)_{\rm cont,\,non}$ 
(i.e.\ $x(e^u+e^d)(x)_{\rm cont}$ at negative $x$) 
obtained from the sum over non-occupied states without smearing, 
for $\gamma\to0$ in (\ref{Eq:smearing}).
The discountinuous nature of the curve is apparent (we connected the points to guide
the eye). It is remarkable that our results practically do not depend on the
intermediate cutoff $\Lambda$. In other words, we observe a very fast 
convergence in $\Lambda$. It is clear that whatever procedure we use to ``smooth'' 
the curve, it will introduce larger numerical uncertainties then the extrapolation
in $\Lambda$, and therefore refrained from performing the latter.
The solid curve in Fig.~\ref{Fig5:x(eu+ed)-unsmeared-calculation}a shows the
final, smoothened curve obtained from fitting the discontinuous curves.

Fig.~\ref{Fig5:x(eu+ed)-unsmeared-calculation}b shows the corresponding results 
for $x(e^u+e^d)(x)_{\rm cont,\,occ}$ obtained from the sum over occupied states.
Also here we observe a stable convergence in 
$\Lambda$, except for very small $x\sim 0.05$,
where anyway the finite box method is not reliable, and the large-$N_c$ 
approach not applicable \cite{Diakonov:1996sr,Diakonov:1997vc}.
The solid curve in Fig.~\ref{Fig5:x(eu+ed)-unsmeared-calculation}b shows the
final result obtained from fitting the discontinuous curves.
(Since the continuum contribution to the quark distribution is far smaller than 
to the antiquark distribution, we computed it with a more coarse-grained resolution 
in $x$.)

What about the respective slow-convergence-$x$-regions?
Here, without smearing, one observes fluctuations similar to that in 
Fig.~\ref{Fig5:x(eu+ed)-unsmeared-calculation}, but two orders of magnitude 
larger, which are due to the anomalous terms, see the discussion above. 

Having switched of  smearing, we could have equally computed $(e^u+e^d)(x)$
(since now computing $x(e^u+e^d)(x)$ or computing $(e^u+e^d)(x)$ 
and multiplying by $x$, of course, commutes).
Still it is preferable to calculate $x(e^u+e^d)(x)$, and the reason
is evident from Fig.~\ref{Fig5:x(eu+ed)-unsmeared-calculation}.
With increasing $x$ the $\Theta$-functions in (\ref{Eq:sperical-symm})
allow to include more discrete states, 
and the fluctuations due to the discrete basis diminish. And vice versa, at small 
$x$ only few long-wave-length states contribute, which enhances the fluctuations 
(and is the reason why here one is particularly sensitive
to details of the finite box method, as mentioned above).
Weighting the function by $x$ suppresses the fluctuations in the small-$x$
region yielding a less fluctuating curve which can be ``smoothened'' more 
reliably, as is demonstrated by Fig.~\ref{Fig5:x(eu+ed)-unsmeared-calculation}.

%------ BEGIN FIGURE 5: unsmeared calculation for x(eu+ed)(x) ------------
%
\begin{figure*}[b]
\begin{tabular}{ccc}
\hspace{-0.5cm}
\includegraphics[height=4.6cm]{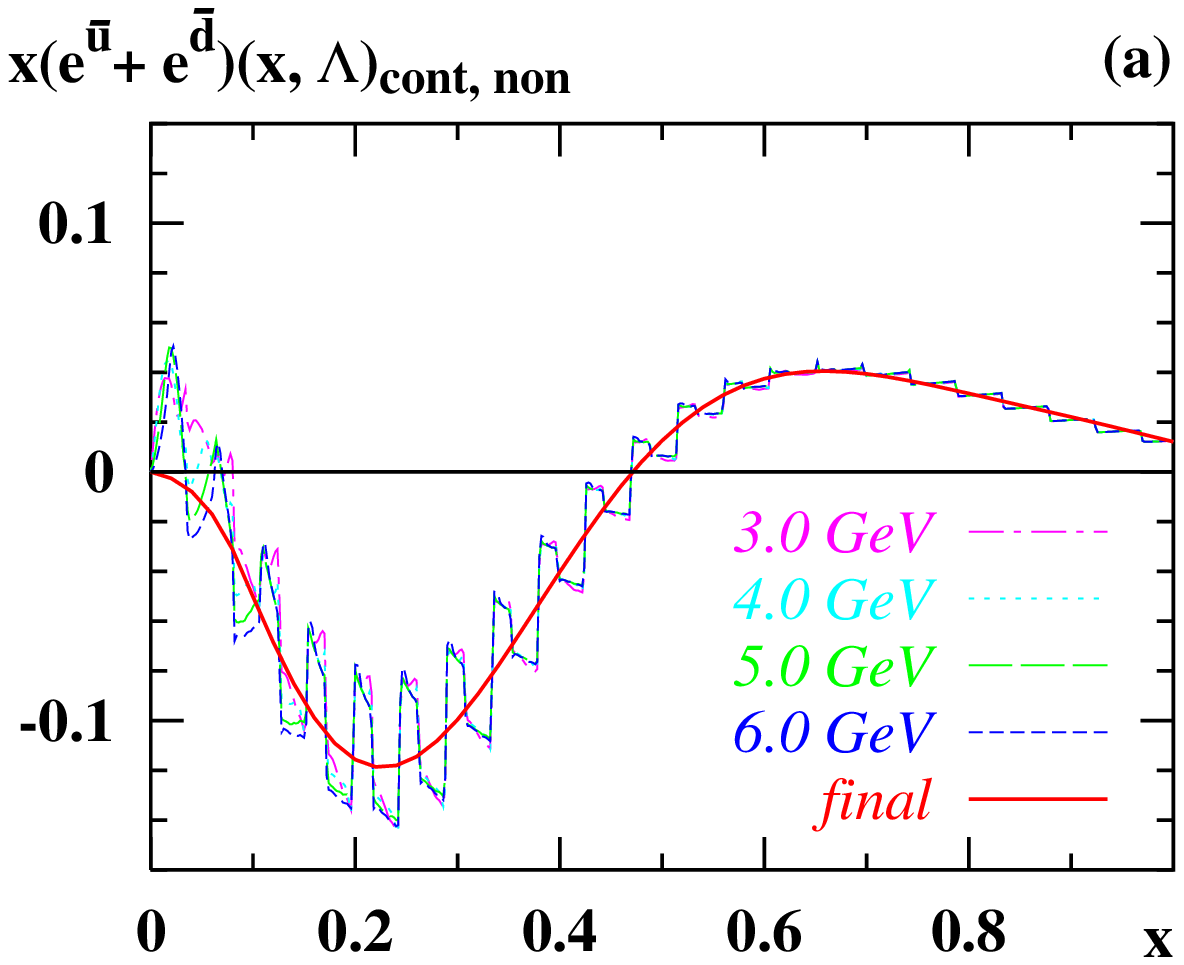} &
\includegraphics[height=4.6cm]{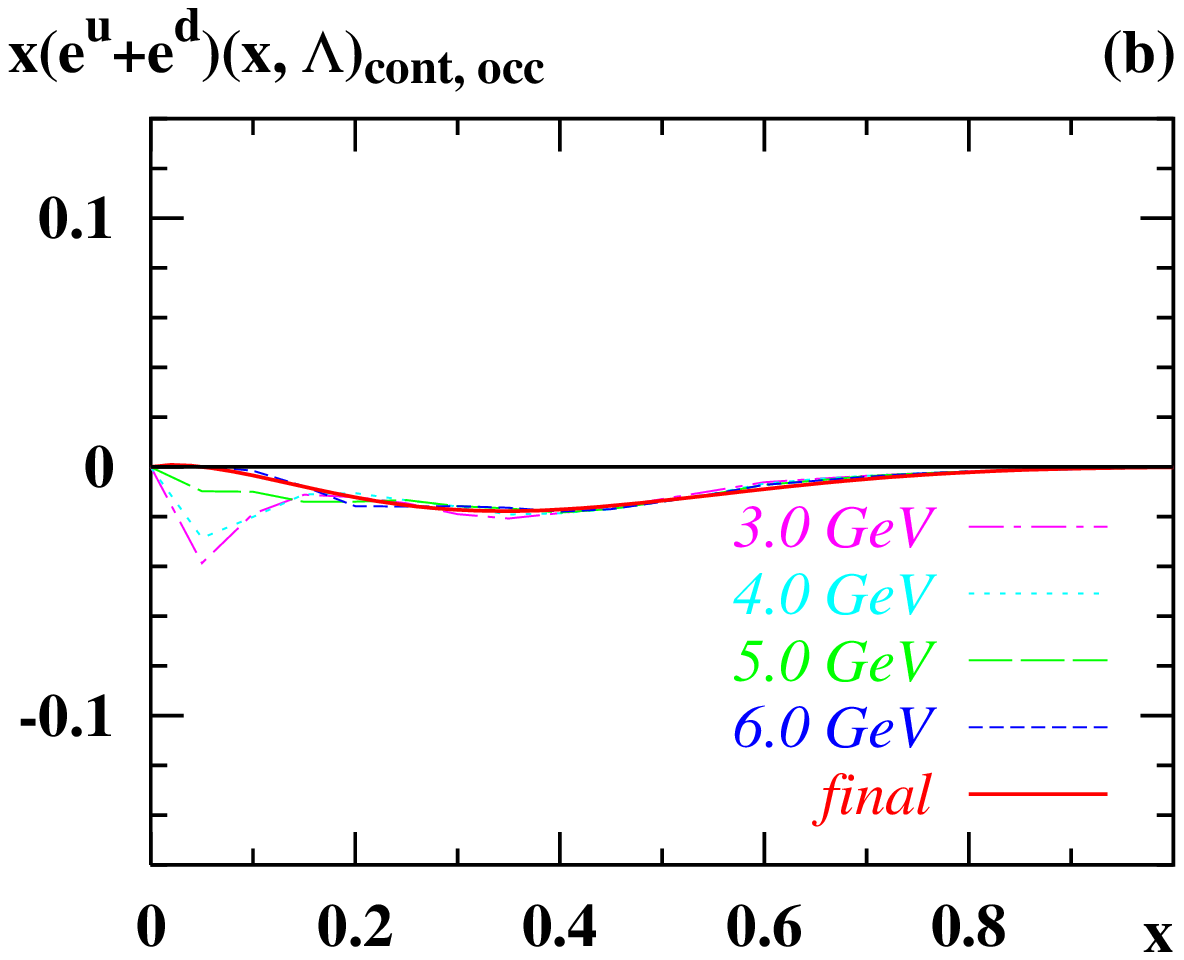} &
\includegraphics[height=4.6cm]{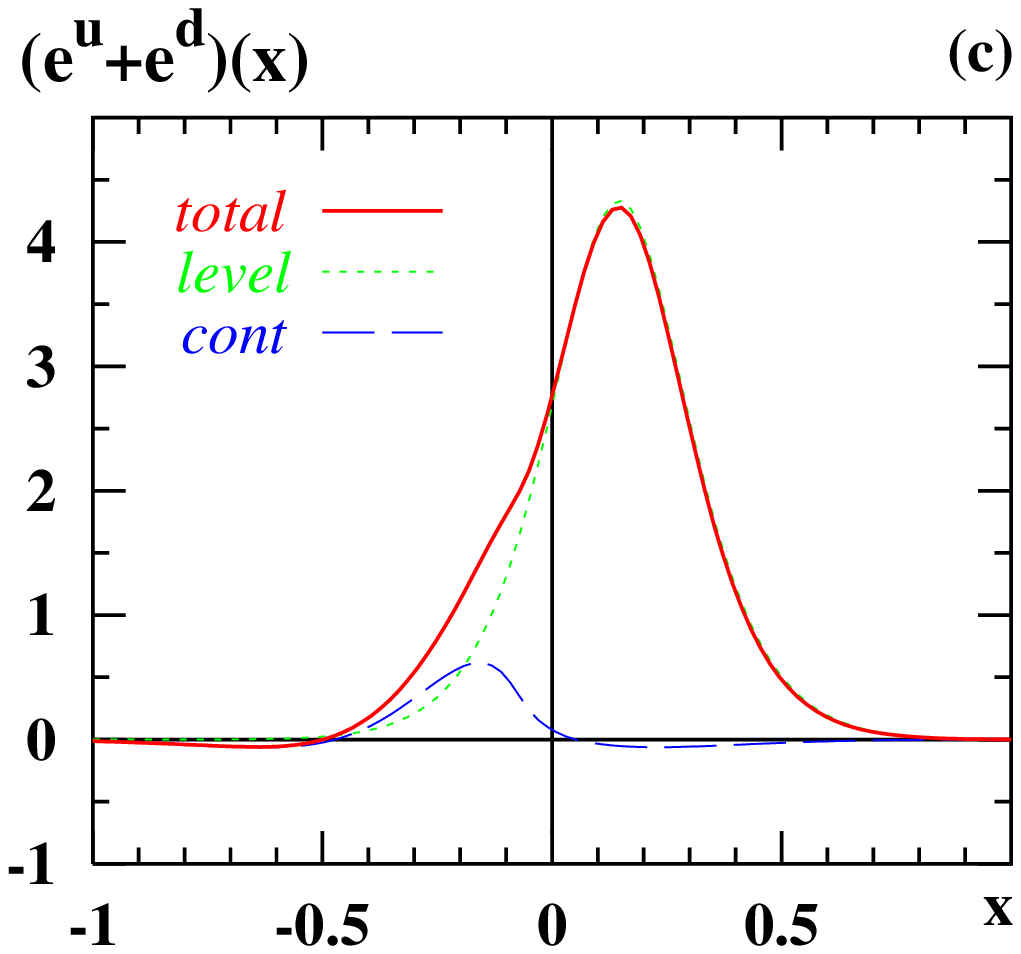} 
\end{tabular}
	\caption{\label{Fig5:x(eu+ed)-unsmeared-calculation}
	The regular continuum contributions to 
	(a) $x(e^{\bar u}+e^{\bar d})(x)$, and
	(b) $x(e^u+e^d)(x)$ as functions of $x$ 
	for the intermediate cutoffs $\Lambda=3,\,4,\,5$ and $6\,{\rm GeV}$,
	and the respective final, smoothened results.
	(c) The total result for the regular part of $(e^u+e^d)(x)$ 
	as function of $x$, and how it is decomposed of the discrete
	level and continuum contributions.}
\end{figure*}
%
%------ END FIGURE 5 -----------------------------------------------------

The final result for the regular part of $(e^u+e^d)(x)$ 
is shown in Fig.~\ref{Fig5:x(eu+ed)-unsmeared-calculation}c.
For completeness we show how it is composed of the contributions from the 
discrete level, Fig.~\ref{fig01:eu+ed-lev}, and the continuum contribution,
Figs.~\ref{Fig5:x(eu+ed)-unsmeared-calculation}a and b.
(In the smoothening step yielding the final (solid) curves in 
Figs.~\ref{Fig5:x(eu+ed)-unsmeared-calculation}a and b we have build in the
constrain that the regular part to the continuum contribution behaves as 
$(e^u+e^d)(x)_{\rm cont}\to {\rm const}\,$ for $x\to0$.) 

As can be seen from Fig.~\ref{Fig5:x(eu+ed)-unsmeared-calculation}c,
the regular continuum contribution to $(e^u+e^d)(x)$ is small compared
to the discrete level contribution which dominates the final result.
Although we could not check the equivalence of the regular 
continuum results from summations over occupied and non-occupied states, 
we still were able to clearly demonstrate that the numerical calculation 
is well under control, which gives confidence into the final results.

%====== SECTION VII: (eu+ed)(x) ======================================
\section{\boldmath Calculation of $(e^u-e^d)(x)$}
\label{Sec-7:(eu-ed)}

%------ BEGIN FIGURE 6: (eu-ed)(x) --------------------------------------
%
\begin{figure*}[t]
\begin{tabular}{cc}
\hspace{-0.5cm}
\includegraphics[height=4.6cm]{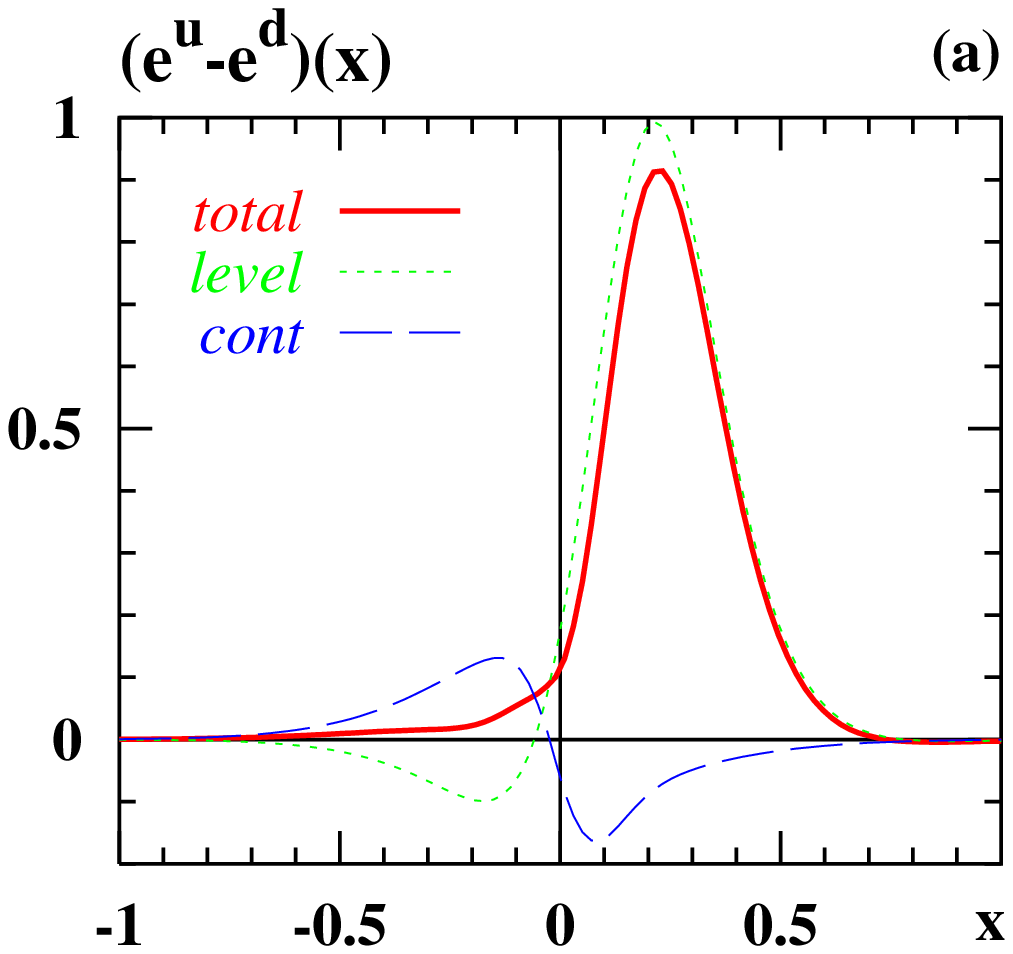}
\includegraphics[height=4.6cm]{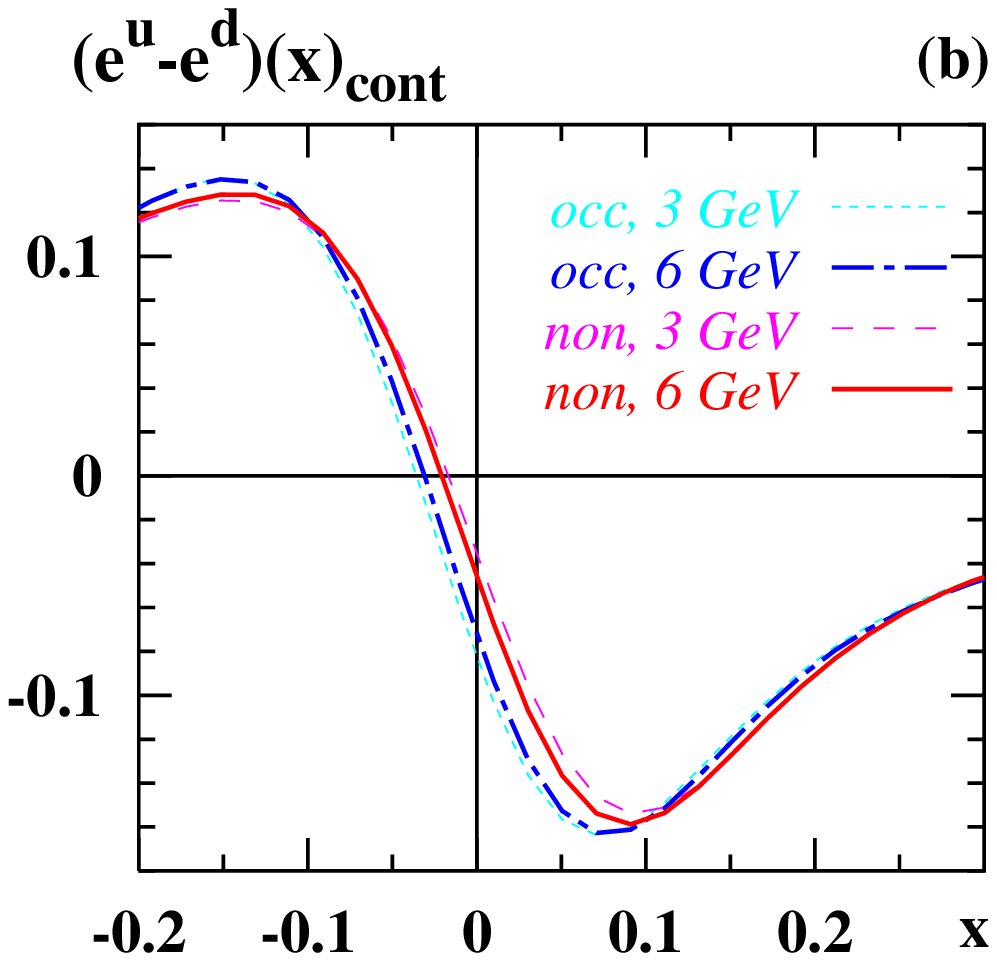} 
\end{tabular}
	\caption{\label{Fig6:eu-ed}
	(a) The total result for $(e^u-e^d)(x)$ as function of $x$, and how 
	it is decomposed from the discrete level and continuum contributions.
	(b) A detail on how the final result for the continuum contribution 
	in Fig.~\ref{Fig6:eu-ed}a comes about: 
	$(e^u-e^d)(x,\Lambda)_{\rm cont}$ as function of $x$ for different 
	intermediate cutoffs from sums over respectively the occupied and 
	non-occupied states.}
\end{figure*}
%
%------ END FIGURE 6 -----------------------------------------------------

The flavour non-singlet combination $(e^u-e^d)(x)$ appears only at subleading 
order in the $1/N_c$ expansion, when one includes ``rotational'' corrections. 
This flavour combination is UV-finite and does not need to be regularized. 
Fig.~\ref{Fig6:eu-ed}a shows the final result for $(e^u-e^d)(x)$.
It is also shown how the total result is decomposed of respectively
the discrete level and continuum contributions.

We observe a fast and stable convergence of the continuum 
contribution $(e^u-e^d)(x,\Lambda)_{\rm cont}$ with increasing intermediate 
cutoff $\Lambda$.
This is demonstrated in Fig.~\ref{Fig6:eu-ed}b where only the region of
strongest  $\Lambda$-dependence around $|x|<0.3$ is shown.
Notice that the different curves in Fig.~\ref{Fig6:eu-ed}b would be nearly
indistinguishable on the scale of Fig.~\ref{Fig6:eu-ed}a.
After the extrapolation $\Lambda\to\infty$ one obtains from the sums over
respectively over occupied and non-occupied states,
Eqs.~(\ref{Eq:model-eu-ed-occ}) and (\ref{Eq:model-eu-ed-non}),
results which coincide to within an accuracy of about $1\%$.

The final result for $(e^u-e^d)(x)$ shown in Fig.~\ref{Fig6:eu-ed} reveals
that it is a regular function of $x$. In particular, no $\delta(x)$-type
singularity appears in this flavour combination.

In order to separate consistently different flavours, $e^u(x)$ and $e^d(x)$, it would
be necessary to consider also rotational corrections to the leading large-$N_c$ 
structure $(e^u+e^d)(x)$. Applying straight-forwardly the procedure 
described in Sec.~\ref{Sec-3:model-and-e} which lead to the expressions
(\ref{Eq:model-eu+ed-occ}-\ref{Eq:model-eu-ed-non}) one obtains rotational 
corrections to $(e^u+e^d)(x)$ consisting only of incomplete double sums. 
This is a general feature encountered whenever one considers $1/N_c$-corrections 
in the model to those parton distribution functions which appear already at leading 
order of the large-$N_c$ expansion \cite{Schweitzer:2001sr}.
Below, when discussing the final results for $e^a(x)$, we shall follow the 
suggestion \cite{Wakamatsu:1998rx,Ohnishi:2003mf} to discard such terms.

%====== SECTION VII: RESULTS & SUM RULES =============================
\section{Discussion of results for \boldmath $e^a(x)$ and sum rules}
\label{Sec-8:results-sum-rules}

\begin{figure*}[t]
\begin{tabular}{ccc}
\hspace{-0.5cm}
\includegraphics[height=5cm]{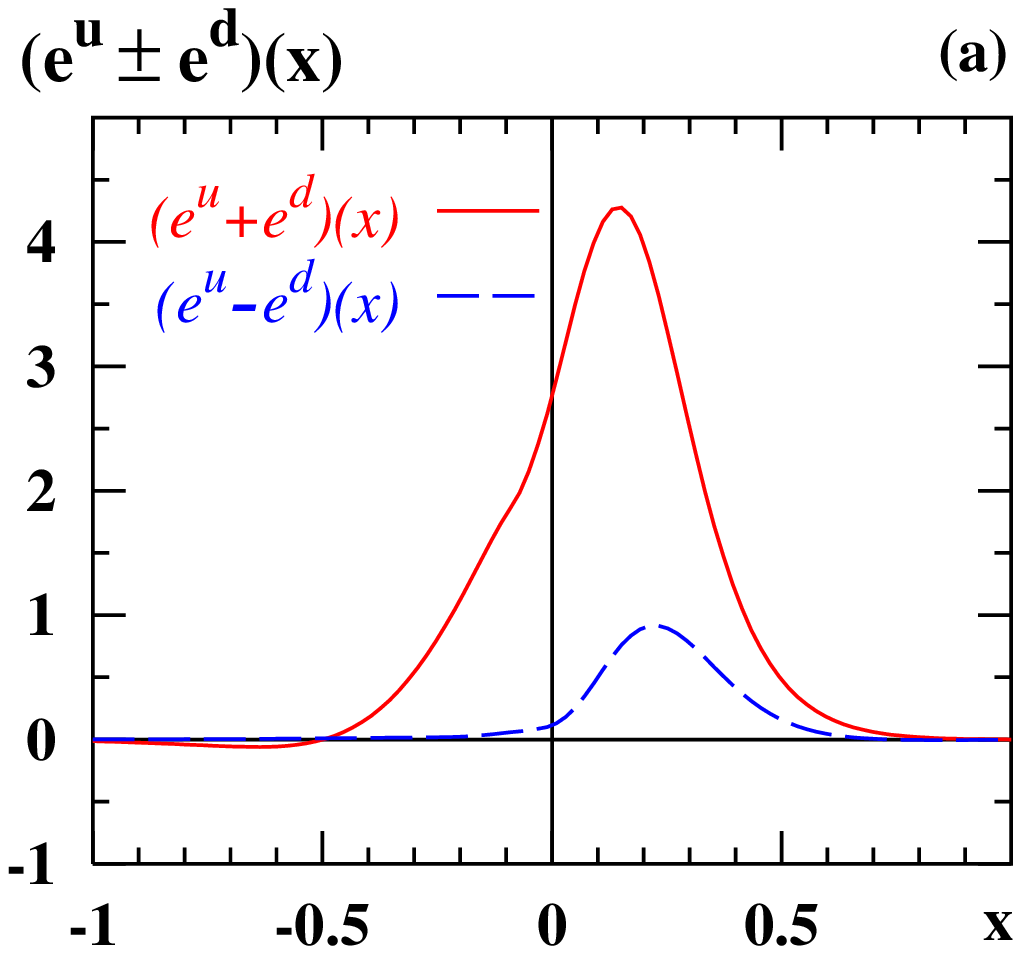} &
\includegraphics[height=5cm]{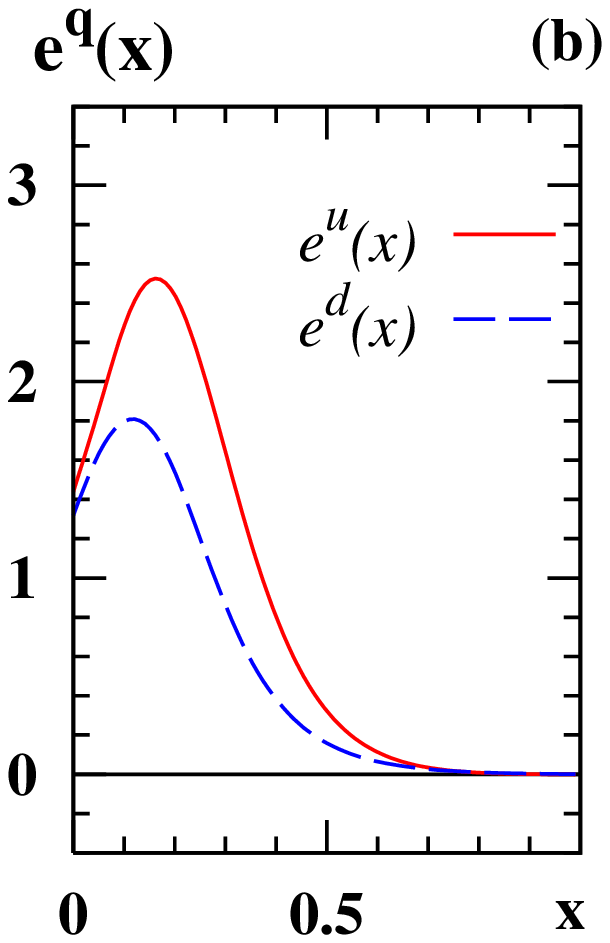} &
\includegraphics[height=5cm]{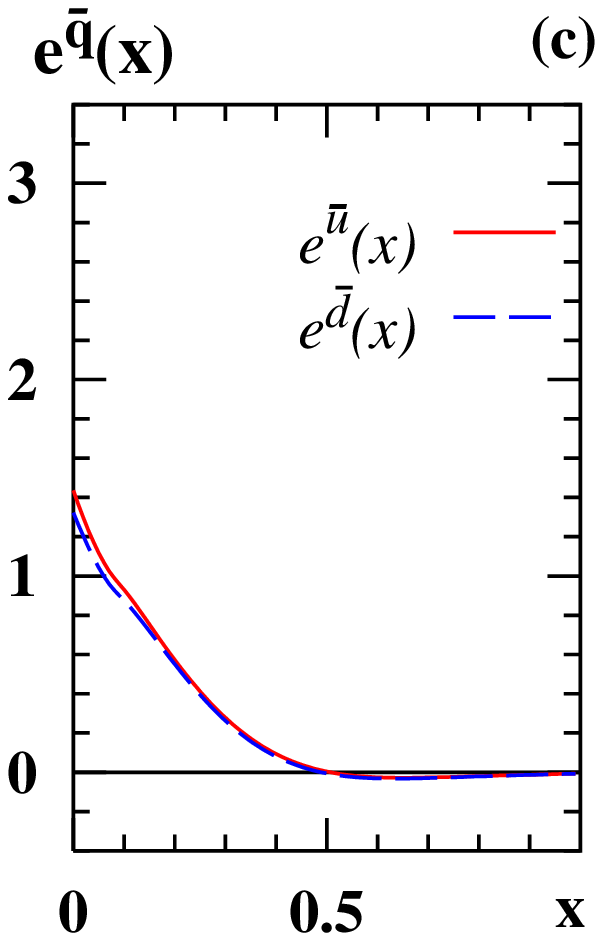}
\end{tabular}
	\caption{\label{Fig7:e-details}
	(a) Comparison of $(e^u+e^d)(x)$ and $(e^u-e^d)(x)$ as functions of $x$.
	(No attempt is made to indicate the $\delta(x)$-contribution in the flavour
	singlet case.)
	(b) The quark distributions $e^u(x)$ and $e^d(x)$.
	(c) The antiquark distributions $e^{\bar u}(x)$ and $e^{\bar d}(x)$.}
%\end{figure*}
%\begin{figure*}[b]
\begin{tabular}{cccc}
\includegraphics[height=5.6cm]{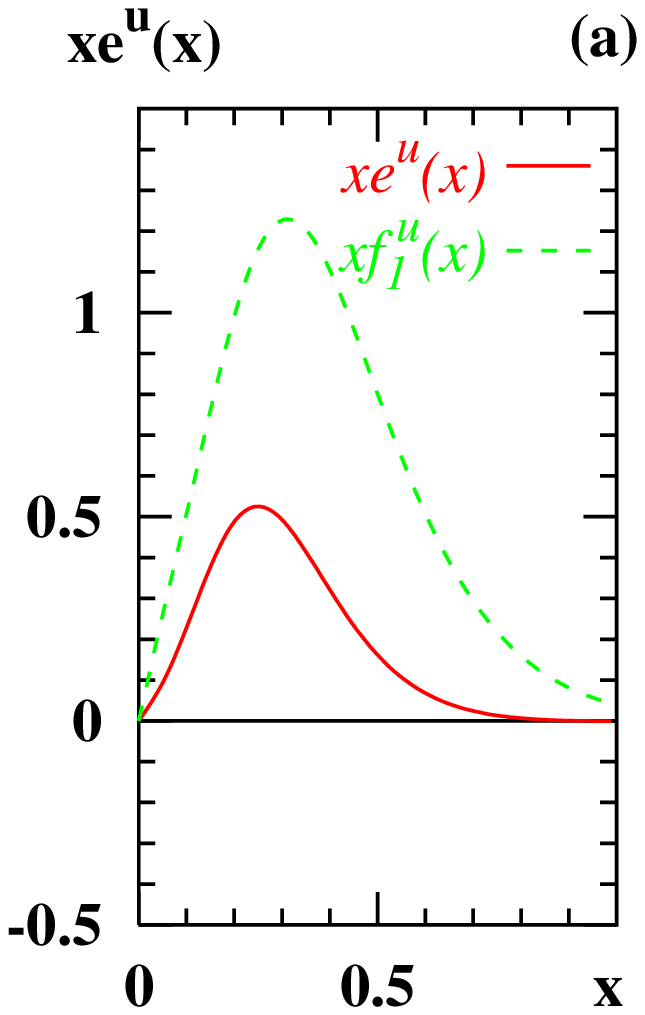} &
\includegraphics[height=5.6cm]{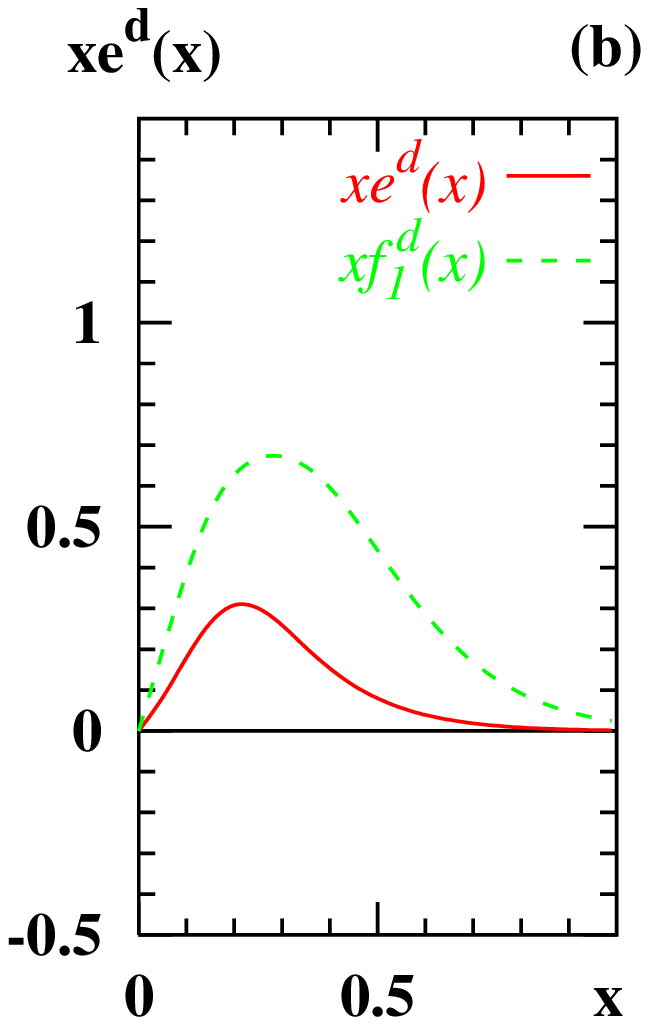} &
\includegraphics[height=5.6cm]{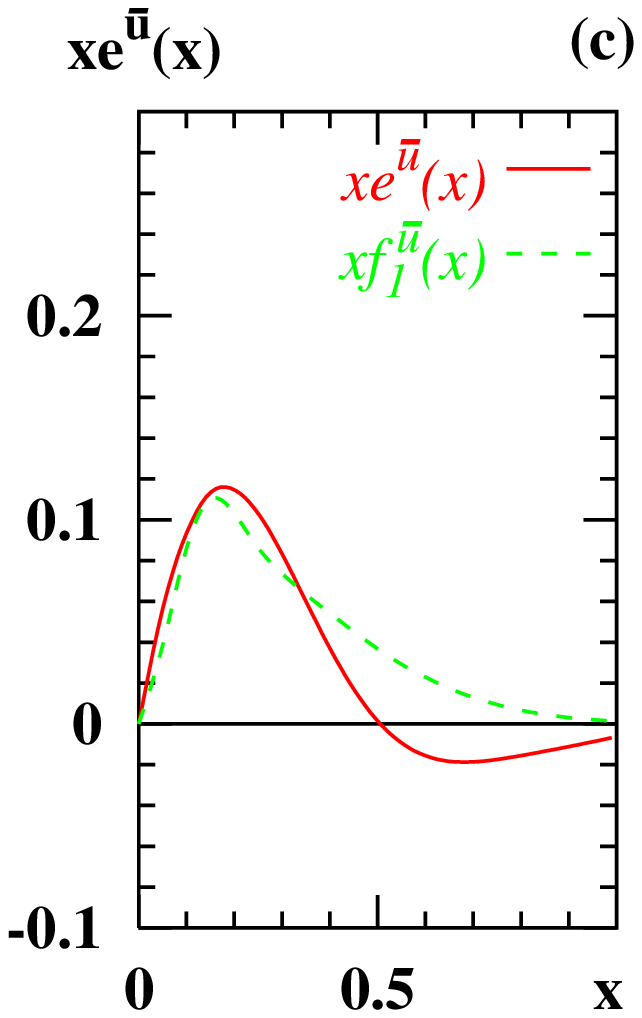} &
\includegraphics[height=5.6cm]{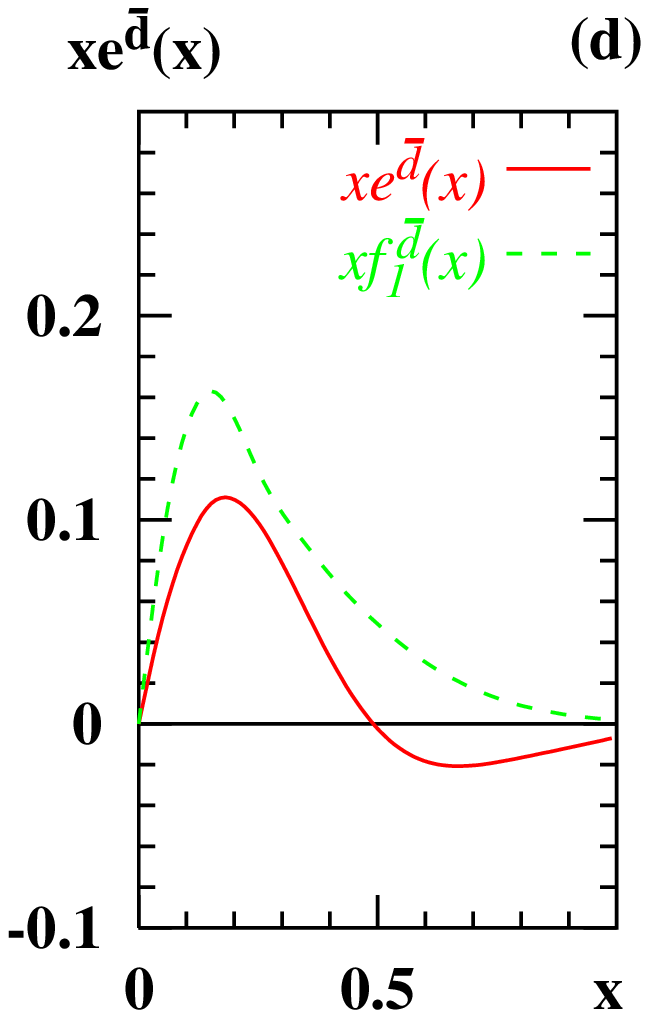} 
\end{tabular}
	\caption{\label{Fig8:xe(x)-vs-xf1(x)}
	The different flavours of $xe^a(x)$ as functions of $x$
	in comparison to $xf_1^a(x)$ from the same model,
	Refs.~\cite{Weiss:1997rt,Pobylitsa:1998tk}.}
\end{figure*}

It is interesting to observe that $e^a(x)$ clearly respects 
the large-$N_c$ predictions for the flavour dependence, Eq.~(\ref{Eq:large-Nc}).
The ``large''  flavour combination $(e^u+e^d)(x)$ is indeed much larger
than the ``small'' flavour combination $(e^u-e^d)(x)$, see 
Fig.~\ref{Fig7:e-details}a.
As a consequence one finds $e^u(x) \approx e^d(x)$ within an accuracy
of about $30\,\%$, see Fig.~\ref{Fig7:e-details}b, which is precisely 
what one generically expects from next-to-leading order corrections 
in an $1/N_c$-expansion with $N_c=3$.
The large-$N_c$ predictions are even more convincingly realized
in the case of antiquarks, see Fig.~\ref{Fig7:e-details}c.

In order to gain some more intuition on the model results for $e^a(x)$
it is instructive to compare them to $f_1^a(x)$ computed in the same model
\cite{Weiss:1997rt,Pobylitsa:1998tk}. The $\chi$QSM results for $f_1^a(x)$ agree to 
within an accuracy of about $30\%$ with parameterizations performed at comparably
low scales \cite{GRV,GRSV}.

Figs.~\ref{Fig8:xe(x)-vs-xf1(x)}a-d show $xe^a(x)$ in comparison
to $xf_1^a(x)$ for $a=u,\,d,\,\bar u,\,\bar d$ at the low scale of the model.
(In this figure it can be seen best that in the model parton distribution 
functions have a non-zero support also for $x>1$ where, however, they are 
proportional to $\exp(-{\rm const}\,N_c x)$ \cite{Diakonov:1996sr}.
Since our results refer to the large-$N_c$ limit, there is conceptually
no problem. In practice, even for $N_c=3$ the distribution functions
are negligibly small for $x>1$, see Fig.~\ref{Fig8:xe(x)-vs-xf1(x)}.)

It is remarkable that the $e^a(x)$ are sizeable only for $x\lesssim 0.5$,
in contrast to the $f_1^a(x)$ which extend also to larger values of $x$.
In the region of $x\lesssim 0.5$ the quark distributions $e^q(x)$ 
are about half the magnitude of the $f_1^q(x)$. However, 
the antiquark distributions $e^{\bar q}(x)$ and $f_1^{\bar q}(x)$ 
are of comparable magnitude.

This comparison is interesting also because it was shown that in the 
non-relativistic limit $e^q(x)$ and $f_1^q(x)$ coincide 
\cite{Efremov:2002qh,Schweitzer:2003uy}.
In the $\chi$QSM, which is a relativistic model, one is far away 
from a non-relativistic scenario, see Figs.~\ref{Fig8:xe(x)-vs-xf1(x)}a~and~b.

%------ BEGIN FIGURE 7+8: x*e^a(x) vs x*f1^a(x) ---------------------------
%
%
%------ END FIGURE 7+8 ---------------------------------------------------

Next let us discuss the sum rules (\ref{e-1moment-a},~\ref{e-1moment-c})
which were analytically proven to be satisfied in the model in
Refs.~\cite{Schweitzer:2003uy,Wakamatsu:2003uu,Ohnishi:2003mf}.
For the first moments we obtain here
\ba
    &&	\int_{-1}^1\!\di x\;(e^u+e^d)(x)
	= 1.76_{\rm lev}
	+ 0.11_{\rm cont,\,reg}
	+ (4.88+0.196)_{\rm cont,\,sing}
	= 6.95\,, 	\label{Eq:mom1-u+d-number}\\
    &&	\int_{-1}^1\!\di x\;(e^u-e^d)(x)
	= 0.303_{\rm lev}
	+ 0.001_{\rm cont}
	= 0.30\,, 	\label{Eq:mom1-u-d-number}
\ea
where we indicate how the numbers are composed of respectively the discrete 
level, and the regular and singular continuum contributions (the last corrected 
for the finite box size effects we know, see Sec.~\ref{Sec-5:(eu+ed)-and-delta}).
The results are in agreement with phenomenology, 
see (\ref{e-1moment-a},~\ref{e-1moment-c}). 
Notice, that the result in (\ref{Eq:mom1-u+d-number}) is strongly sensitive 
to details of regularization. With the exact result for the coefficient $C$ we 
obtain $\int\di x\,(e^u+e^d)(x)\simeq 12$, see Sec.~\ref{Sec-5:(eu+ed)-and-delta}.

While the sum rules (\ref{e-1moment-a},~\ref{e-1moment-c}) are formally satisfied
in the model \cite{Schweitzer:2003uy,Wakamatsu:2003uu,Ohnishi:2003mf} and 
numerically in satisfactory agreement with phenomenology one must admit a shortcoming.
In QCD the sum rules (\ref{e-1moment-a},~\ref{e-1moment-c}) are saturated 
{\sl solely} by the $\delta(x)$-contribution, see Sec.~\ref{Sec-2:e-theory}.
In contrast to that in the model about $20\,\%$ of the result in 
(\ref{Eq:mom1-u+d-number}) are due to the regular part of 
$(e^u+e^d)(x)$, while the total result in (\ref{Eq:mom1-u-d-number})
is due to the regular (and only) part of $(e^u-e^d)(x)$.

Another shortcoming is that in the chiral limit to which our results refer 
the second moments (\ref{e-2moment}) for $(e^u\pm e^d)(x)$ vanish in QCD.
Instead, we obtain in the model
\ba
    &&	\int_{-1}^1\!\di x\;x(e^u+e^d)(x)
	= 0.258_{\rm lev}
	- 0.024_{\rm cont}
	= 0.23 \,, 	\label{Eq:mom2-u+d-number}\\
    &&	\int_{-1}^1\!\di x\;x(e^u-e^d)(x)
	= 0.089_{\rm lev}
	- 0.019_{\rm cont}
	= 0.07 \,.	\label{Eq:mom2-u-d-number}
\ea
These two shortcomings have the same origin. 

Both, the fact that the sum rule 
(\ref{e-1moment}) is saturated solely by the $\delta(x)$-contribution and 
the sum rule (\ref{e-2moment}), follow from applying explicitly the QCD equations 
of motion. At this point the $\chi$QSM and actually any model is overburdened. 
Effective model approaches do not respect the QCD equations of motion, 
rather they satisfy the respective model equations of motion.
Therefore it is not surprizing to observe such sum rules not to be satisfied 
literally. Still, one may explore model equations of motions and reinterpret
the sum rules  (\ref{e-1moment},~\ref{e-2moment}) in the model terminology
\cite{Schweitzer:2003uy,Ohnishi:2003mf}.

%====== SECTION IX: COMPARE TO PREVOUS CALCULATIONS IN CQSM ==========
\section{Comparison to previous calculations in the \boldmath $\chi$QSM}
\label{Sec-9:previous-in-model}

The first (approximate) calculation of $e^a(x)$ in the $\chi$QSM was reported in 
\cite{Schweitzer:2003uy}.
There the regular part of $(e^u+e^d)(x)$ was approximated by the discrete level 
contribution, and rotational corrections were discarded. The accuracy of these
approximations was estimated to be within $30\,\%$. 
In fact, the (regular) continuum contributions are small 
compared to the respective discrete level contributions, see 
Figs.~\ref{Fig5:x(eu+ed)-unsmeared-calculation}c and \ref{Fig6:eu-ed}a, 
and the rotational correction $(e^u-e^d)(x)\ll(e^u+e^d)(x)$, see 
Sec.~\ref{Sec-8:results-sum-rules}.
Thus, the estimates of \cite{Schweitzer:2003uy} did indeed approximate $e^a(x)$ 
within the claimed accuracy. Of course, this became clear only after exact 
calculations were presented.

The first exact calculation of $e^a(x)$ in the $\chi$QSM was presented in
\cite{Ohnishi:2003mf}, and the present work confirms the results obtained there. 
The main quantitative difference between our work and \cite{Ohnishi:2003mf} 
is that there it was possible to handle much larger Pauli-Villars masses
(which requires, see Sec.~\ref{Sec-4:(eu+ed)}, more computing power 
available in \cite{Ohnishi:2003mf}).

The choice of Pauli-Villars masses is relevant for the continuum contribution 
$(e^u+e^d)(x)$. As a consequence of the larger  Pauli-Villars masses in 
\cite{Ohnishi:2003mf} an about two times larger coefficient of the 
$\delta(x)$-contribution was obtained. 
But the results for the regular contribution to $e^a(x)$
obtained here and in \cite{Ohnishi:2003mf} practically agree.
In fact, these results are strongly dominated by the discrete level contributions, 
see Figs.~\ref{Fig5:x(eu+ed)-unsmeared-calculation}c and \ref{Fig6:eu-ed}a, and at
this point the slightly different value of the constituent mass $M=375\,{\rm MeV}$
used in \cite{Ohnishi:2003mf} (compared to $M=350\,{\rm MeV}$ used here) is more
decisive than the different choice of Pauli-Villars masses.

The present work extends in several aspects Ref.~\cite{Ohnishi:2003mf}, which was 
up to now the most detailed and complete study of $e^a(x)$ in this model. 
\begin{itemize}
\item	We have shown that the coefficient $C$ of the 
	$\delta(x)$-term in $(e^u+e^d)(x)$ can be determined {\sl exactly} 
	by means of an analytical calculation \cite{Schweitzer:2003uy}.
	% This is a unique case in the $\chi$QSM that it is possible to compute exactly
	% a contribution (albeit a rather particular and unusual one) to a 
	% parton distribution function.
\item 	We have determined finite box size corrections 
	for the coefficient $C$  of the $\delta(x)$-term in $(e^u+e^d)(x)$.
\item 	We demonstrated the equivalence of results obtained from summations
	over occupied and non-occupied states, wherever possible.
	(In \cite{Ohnishi:2003mf} the equivalence was demonstrated only for 
	the coefficient $C$.)
\item 	Where this was not possible, namely for the regular continuum part of 
	$(e^u+e^d)(x)$, we were able to explain why, and to demonstrate that 
	nevertheless the involved numerics is under analytical control.
\end{itemize}

In view of the complexity of the task --- to deal numerically with a 
$\delta(x)$-term, to use a double Pauli-Villars subtraction with large 
Pauli-Villars masses --- the present work provides an important 
supplement to Refs.~\cite{Schweitzer:2003uy,Wakamatsu:2003uu,Ohnishi:2003mf}.

%====== SECTION X: COMPARE TO OTHER MODELS ===========================
\section{Comparison to results from other models}
\label{Sec-10:compare-to-other-models}

\begin{figure*}[t]
\begin{tabular}{cccc}
\hspace{-0.5cm}
\includegraphics[height=5cm]{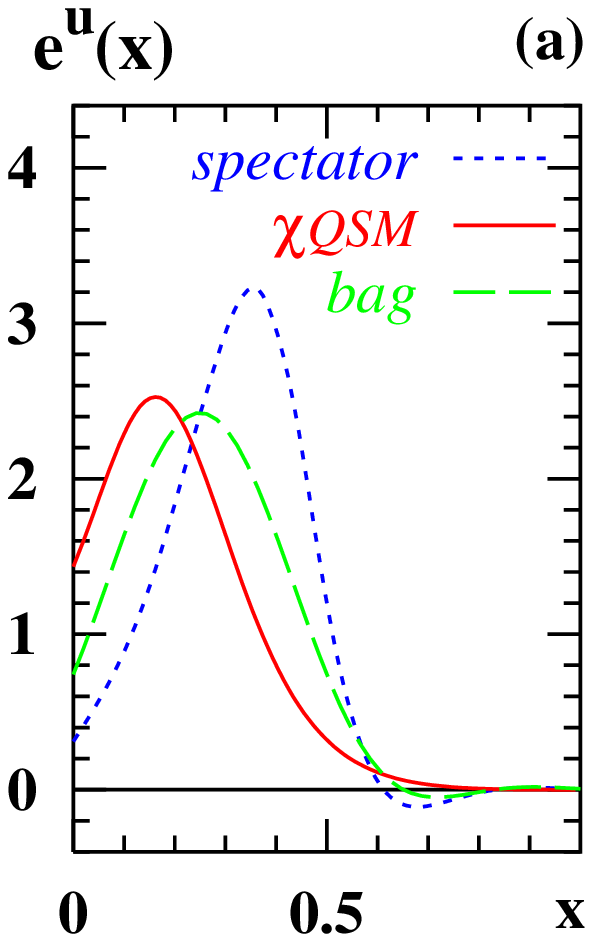}&
\includegraphics[height=5cm]{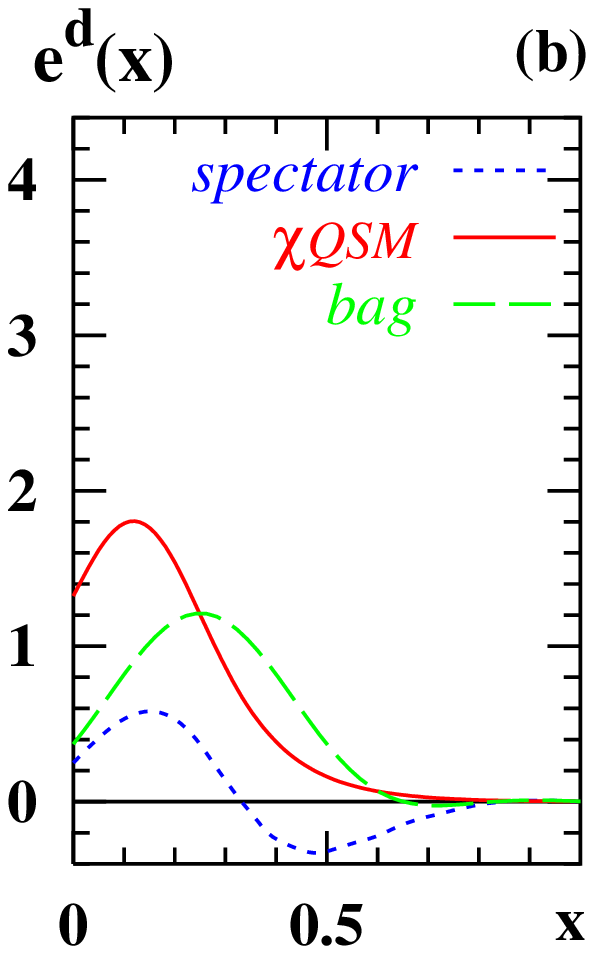}&
\includegraphics[height=5cm]{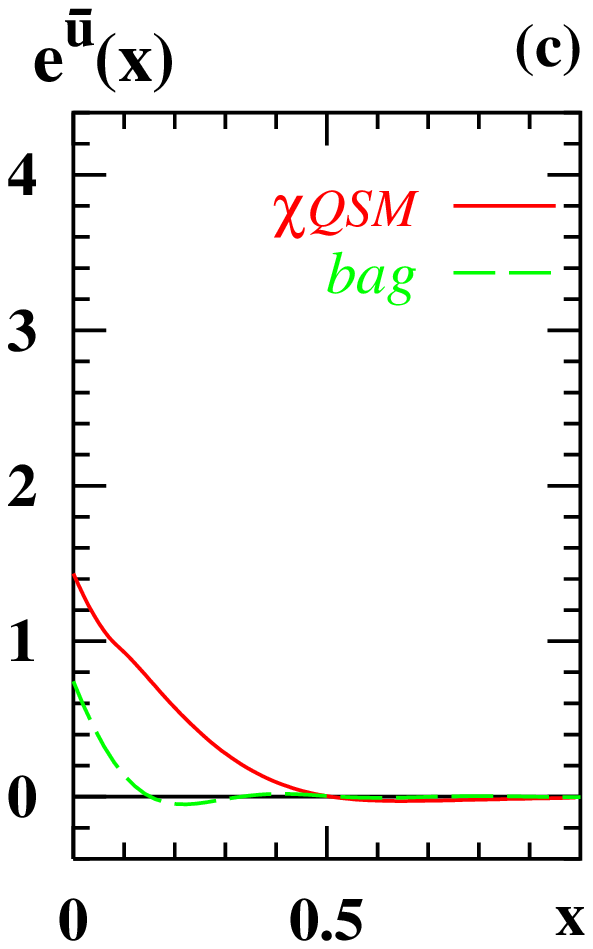}&
\includegraphics[height=5cm]{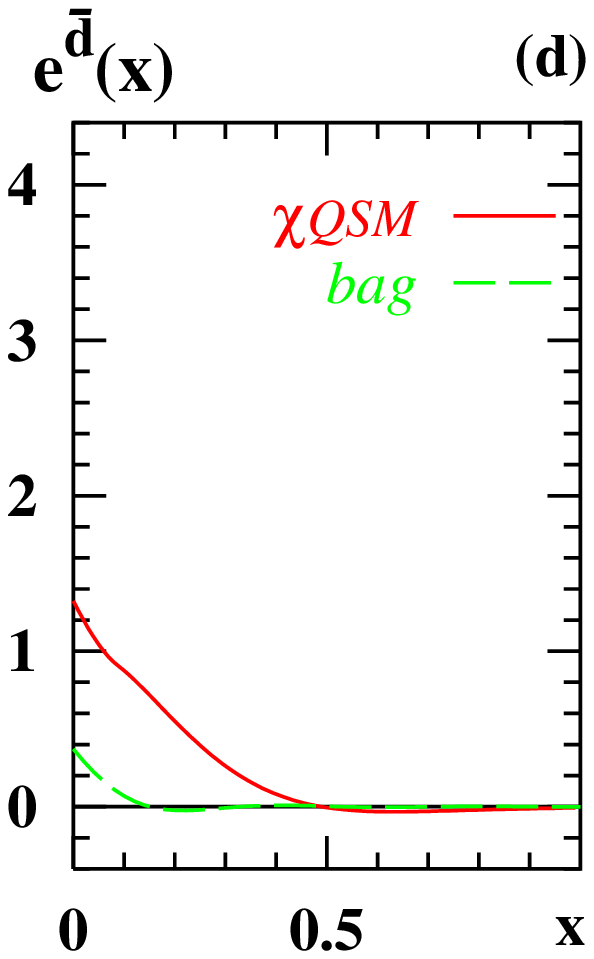}
\end{tabular}
\hspace{-0.5cm}
	\caption{\label{Fig9:e-models}
	Comparison of predictions for $e^a(x)$ from different models.
	Solid line: $\chi$QSM, computed in this work.
	Dashed line: bag model \cite{Jaffe:1991ra}.
	Dotted line: spectator model \cite{Jakob:1997wg}.}
\end{figure*}

The distribution function $e^a(x)$ was studied also in other non-perturbative 
model-approaches, such as bag \cite{Jaffe:1991ra,Signal:1996ct} or spectator 
\cite{Jakob:1997wg} models, as well as in 1+1 dimensional toy-models or 
perturbative one-loop model calculations \cite{Burkardt:1995ts,Burkardt:2001iy}.
In Fig.~\ref{Fig9:e-models} we compare our results to the MIT bag model 
calculation \cite{Jaffe:1991ra} and the spectator model \cite{Jakob:1997wg}.

In the bag model version used in \cite{Jaffe:1991ra} the nucleon is assumed 
to consist of 3 non-interacting, massless quarks confined to the interior 
of a 3D spherical cavity (``bag'') by imposing appropriate covariant 
boundary conditions which model confinement and mimic gluonic effects.
The model is relativistic. The flavour dependence is due to the 
assumed SU(2)$_{\rm flavour}\times$SU(2)$_{\rm spin}$ spin-flavour-symmetry of the 
quark wave functions such that $e^u(x)=2\,{\rm e}(x)$ and $e^d(x)={\rm e}(x)$, and 
analogously for antiquarks, with ${\rm e}(x)$ as introduced in \cite{Jaffe:1991ra}.
There is no $\delta(x)$-contribution in the bag model, and since the quarks 
are massless, the $e^a(x)$ computed in \cite{Jaffe:1991ra} correspond to
$e^a_{\rm tw3}(x)$ in Eq.~(\ref{e-decomposition}).
Interestingly, the pure twist-3 (``interaction-dependent'') nature of this
contribution is reflected in the bag model by the fact that $e^a(x)$ is due 
to bag-boundary effects. The sum rules (\ref{e-1moment},~\ref{e-2moment}) 
are not satisfied in the bag model because the QCD equations of motion are 
modified in the bag \cite{Jaffe:1991ra}, c.f.\ the discussion in the 
$\chi$QSM in Sec.~\ref{Sec-8:results-sum-rules}.

The bag gives also rise to antiquark distributions, however, to unphysical ones
since the unpolarized antiquark distributions in the bag model violate positivity,
i.e.\ in this model $f_1^{\bar q}(x)<0$ is found. Only valence quark distributions are
considered to be physical \cite{Jaffe:1974nj}. Keeping this mind, we plot also 
$e^{\bar q}(x)$ from the bag model in Fig.~\ref{Fig9:e-models}c and d.

In spectator models parton distribution functions are modelled
by introducing a unity in the shape of $\sum_n |n\ra\,\la n|=1$
into the definition, here Eq.~(\ref{Eq:def-e}), where $\{|n\ra\}$
is a complete set of intermediate states, and modelling this complete 
set of states by e.g.\ a diquark state. Hereby the diquark state can be 
taken to be, e.g., in a spin 0 or spin 1 state, which is referred to as
respectively scalar and vector diquark. Both were considered in 
\cite{Jakob:1997wg} which yields, in spite of the 
SU(2)$_{\rm flavour}\times$SU(2)$_{\rm spin}$ spin-flavour-symmetry
assumed also there, to a non-trivial flavour dependence in
Figs.~\ref{Fig9:e-models}a,~b.
Antiquark distributions can be included by introducing more complex 
intermediate spectator states, but were not considered in \cite{Jakob:1997wg}.
Also the spectator model does not respect the sum rules
(\ref{e-1moment},~\ref{e-2moment}).

When comparing the different models of $e^a(x)$ in Fig.~\ref{Fig9:e-models} 
one has to keep in mind that the low scales in the various models differ somehow. 
Bag model results refer to a scale of about $0.4\,{\rm GeV}$,
spectator model results to about $0.5\,{\rm GeV}$, and the scale of 
$\chi$QSM results is roughly $0.6\,{\rm GeV}$, see Eq.~(\ref{scale}).

The three models agree on that $e^u(x)$ is positive and sizeable for $x\lesssim 0.5$, 
see Fig.~\ref{Fig9:e-models}a, while  $e^d(x)$ appears to be smaller 
(and exhibits in the spectator model even a remarkable zero around $x\sim 0.3$), 
see Fig.~\ref{Fig9:e-models}b. 
The $e^{\bar q}(x)$ are much larger in the $\chi$QSM model than in the bag model,
see Figs.~\ref{Fig9:e-models}c and d.
Only in the $\chi$QSM there is a $\delta(x)$-contribution at $x=0$
(no attempt was made to indicate this singularity in Fig.~\ref{Fig9:e-models}).

%------ BEGIN FIGURE 9: x*e^a(x) vs x*f1^a(x) ----------------------------
%
%
%------ END FIGURE 9 -----------------------------------------------------

%\newpage
%====== SECTION XI: SUMMARY & CONCLUSIONS ============================
\section{Conclusions}
\label{Sec-11:conclusions}

A study of the distribution functions $e^a(x)$
in the $\chi$QSM was presented which completes previous works 
\cite{Schweitzer:2003uy,Wakamatsu:2003uu,Ohnishi:2003mf}.
Two particular features not encountered before in 
calculations of other parton distribution functions
complicate the computation of $e^a(x)$  in the $\chi$QSM: the appearances of a 
$\delta(x)$-singularity and of quadratic UV-divergences
whose regularization requires a double Pauli-Villars subtraction with
a large second Pauli-Villars mass $M_2={\cal O}(1\,{\rm GeV})$.
We have demonstrated in detail that in spite of these complications 
a reliable, controlled numerical calculation is possible. 
Our results confirm qualitatively and quantitatively previous calculations of 
$e^a(x)$ in the $\chi$QSM \cite{Schweitzer:2003uy,Wakamatsu:2003uu,Ohnishi:2003mf}.

Moreover, our study reveals several interesting results. 
We have demonstrated that the coefficient $C$ 
of the $\delta(x)$-singularity in $(e^u+e^d)(x)$ can be 
calculated analytically in the model \cite{Schweitzer:2003uy}.
As far we are aware this the only case where it is possible to
compute {\sl exactly} a (though admittedly unusual) contribution to 
a parton distribution function in the $\chi$QSM.
This means that the coefficient $C$ of the $\delta(x)$-term is exactly 
proportional to the quark vacuum condensate, and thus ultimately connected to 
chiral symmetry breaking \cite{Schweitzer:2003uy}.

Another interesting, but more technical byproduct of our study is that we have been 
able to quantify exactly the finite-box-size corrections to a quantity in the model,
namely to the coefficient $C$ (which are of the order of few $\%$).
This is also, to best of our knowledge, unique in the model.

The $\chi$QSM predicts $e^u(x)$ to be positive and sizeable 
--- reaching half the size of $f_1^u(x)$ ---
in the region of $x\lesssim 0.5$, and $e^d(x)$ somehow smaller.
Remarkably, in this region of $x$ the $e^{\bar q}(x)$ appear 
to be of similar magnitude as the $f_1^{\bar q}(x)$.
Our predictions for the quark distributions $e^q(x)$ are in rough 
qualitative agreement with results from bag \cite{Jaffe:1991ra,Signal:1996ct} 
or spectator \cite{Jakob:1997wg} models in which, however, 
no $\delta(x)$-contribution appears.

Here, following the practical point of view of Ref.~\cite{Jaffe:1991ra}, 
we computed $e^a(x)$ benefiting from a special property of this twist-3 
quantity,  which allows to define it in terms of quark fields only, and 
thus makes possible studies in models without explicit gluon degrees 
of freedom \cite{Jaffe:1991ra}. However, by means of QCD equations of motion
$e^a(x)$ can be reexpressed such that they explicitly depend on gluon fields.
The question whether quark models, like the $\chi$QSM, % bag or spectator model, 
are nevertheless able to provide useful estimates for $e^a(x)$,
at least in certain regions of $x$, can be clarified only by experiment.

%\newpage
\begin{acknowledgments}
We thank Klaus Goeke for numerous discussions.
The work is partially supported by BMBF (Verbundforschung), 
and is part of the European Integrated Infrastructure Initiative Hadron
Physics project under contract number RII3-CT-2004-506078. 
D.~U.\  acknowledges support from GRICES and DAAD.
\end{acknowledgments}

%====== REFERENCES ===================================================

\end{document}